\documentclass[conf]{new-aiaa}
\usepackage[utf8]{inputenc}

\usepackage{graphicx}
\usepackage{amsmath}
\usepackage[version=4]{mhchem}
\usepackage{siunitx}
\usepackage{caption}
\usepackage{subcaption}
\usepackage{longtable,tabularx}
\usepackage{setspace}
\usepackage{cancel}
\usepackage{float}
\usepackage{multirow}
\usepackage{multicol}
\usepackage{lineno}

\title{Performance of Flamelet Models with Epsilon Tracking for Diffusion Flame Simulations}

\author{Sylvain L. Walsh\footnote{Ph.D student, Department of Mechanical and Aerospace Engineering, AIAA Member.}, Yalu Zhu\footnote{Assistant Specialist, Department of Mechanical and Aerospace Engineering, AIAA Member.}, Feng Liu\footnote{Professor, Department of Mechanical and Aerospace Engineering, AIAA Fellow.} and William A. Sirignano\footnote{Distinguished Professor, Department of Mechanical and Aerospace Engineering, AIAA Honorary Fellow.}}
\affil{University of California, Irvine, Irvine, CA, 92697}

\begin{document}

\maketitle
\begin{abstract}
This work examines the physical consistency of the conventional Flamelet Progress Variable (FPV) model for diffusion flame simulations and and introduces a new compressible flamelet formulation that employs the turbulent kinetic energy dissipation rate, $\boldsymbol{\epsilon}$, as the tracking variable. Two-dimensional Reynolds-averaged Navier-Stokes (RANS) simulations are conducted for a reacting, transonic, turbulent mixing layer to assess the coupling between resolved-scale and subgrid flamelet quantities, with emphasis on the role of strain rate. The FPV model is found to decouple resolved-scale and subgrid strain rates, leading to the preferential selection of equilibrium flamelet solutions in regions of high strain and resulting in nonphysical predictions of heat release and species composition. The proposed $\boldsymbol{\epsilon}$-based formulation restores physical consistency by relating the subgrid flamelet strain rate to $\boldsymbol{\epsilon}$, allowing the flamelet to respond to the local resolved-scale strain field. The inclusion of resolved-scale species transport enables advective and diffusive redistribution of products across locally quenched regions. The results indicate that $\boldsymbol{\epsilon}$ offers a physically consistent tracking variable that connects the sub-grid flamelet model to resolved-scale RANS computations.
\end{abstract}

\section*{Nomenclature}
{\renewcommand\arraystretch{1.0}
\noindent\begin{longtable*}{@{}l @{\quad=\quad} l@{}}
$C$ & progress variable \\
$C_{\mu}$ & $k-\epsilon$ turbulence model empirical constant\\
$C_{\chi}$ &  proportionality constant between turbulence and scalar time-scales\\
$C_p$ & specific heat capacity at constant pressure \\
$E$ & total sensible energy \\
$h$ & sensible enthalpy\\
$h^0_f$ & enthalpy of formation \\
$\mathbf{I}$ & Kronecker tensor \\
$\mathbf{j}$ & diffusive flux \\
$k$ & turbulent kinetic energy \\
$N$ & total number of species\\
$p$ & pressure \\
$\mathrm{Pr}$ & Prandtl number\\
$\dot{Q}$ & heat release rate\\
$R$ & gas constant \\
$R_0$ & universal gas constant, $R_0$ = 8.3145 J/mol/K \\
$\mathrm{Sc}$ & Schmidt number\\
$T$ & temperature\\
$T_{\mathrm{ref}}$ & reference temperature, $T_{\mathrm{ref}}=298.15$ K \\
$t$ & time\\
$u$ & streamwise velocity\\
$\mathbf{V}$ & velocity vector\\
$W$ & molecular weight \\
$Y$ & mass fraction\\
$Z$ & mixture fraction\\
$Z^{''2}$ & variance of the mixture fraction\\
$\chi$ & instantaneous scalar dissipation rate\\
$\epsilon$ & turbulence kinetic energy dissipation rate\\
$\lambda$ & flamelet parameter \\
$\mu$ & mixture-averaged coefficient of viscosity \\
$\omega$ & specific turbulent dissipation rate \\
$\dot{\omega}$ & species production rate or progress variable \\
$\psi$ & reactive scalar\\
$\rho$ & density \\
$\boldsymbol{\tau}$ & viscous stress tensor \\

\end{longtable*}}
\textit{Subscripts}
{\renewcommand\arraystretch{1.0}
\noindent\begin{longtable*}{@{}l @{\quad=\quad} l@{}}
$n$ & species specific quantity\\
$st$ & stoichiometric composition\\
$T$ & turbulent quantity\\
\end{longtable*}}

\textit{Superscripts}
{\renewcommand\arraystretch{1.0}
\noindent\begin{longtable*}{@{}l @{\quad=\quad} l@{}}
$-$ & Reynolds-averaged quantity\\
$\thicksim$ & Favre-averaged quantity\\
\end{longtable*}}

\section{Introduction}\label{intro}
\lettrine{N}{umerical} simulation of turbulent combustion presents a formidable challenge due to the coupled complexities of turbulence and chemical kinetics. Turbulence introduces a broad spectrum of spatial and temporal scales, along with unresolved scales that give rise to closure and modeling issues. Combustion further complicates the problem through its inherently multi-step nature, involving finite-rate chemistry with possibly hundreds of reactive species. These reactions often occur at time scales smaller than the time scales for the smallest turbulent eddies.

Accurate prediction of chemical heat release necessitates the resolution of detailed finite-rate kinetics, which in turn requires solving a transport equation for each participating species. The permissible numerical timestep is constrained by the fastest chemical timescales, typically associated with short-lived radical species. Although these radicals are minor in terms of composition, they are essential to correctly capturing reaction physics. The stiffness introduced by such species, combined with the large number of transport equations required, renders direct numerical simulation (DNS) computationally intractable for practical configurations. Consequently, most simulations rely on modeling approaches, such as large eddy simulation (LES) or Reynolds-averaged Navier–Stokes (RANS) formulations. The use of such approaches necessitates modeling of unresolved chemical terms, which remains a major area of ongoing research. As a result of these challenges, the current application of computational modeling for thermal systems in engineering practice is often limited. Confidence in the accurate prediction of finite-rate chemistry effects remains low, leading to their occasional use and typically only marginal influence on the design and development process \cite{bilger_role_2011}.

A straightforward approach to combustion modeling is to compute chemical source terms directly at the resolved scale, using Arrhenius-type rate expressions based on local species concentrations and thermodynamic conditions. This method becomes computationally expensive in LES and RANS due to the large number of species transport equations and small timestep limitations in detailed multi-step kinetics. As a result, the chemistry is often reduced to a one-step kinetics (OSK) global mechanism or a small set of representative reactions.
This simplification introduces two key issues. First, the loss of intermediate species and reaction pathways leads to inaccurate chemical source terms. Second, relying on resolved quantities to compute chemical rates overlooks the fact that combustion occurs on scales smaller than those resolved by LES or RANS. This motivates the use of reduced order subgrid-scale combustion models that can incorporate detailed finite-rate kinetics while accounting for turbulence–chemistry interactions.

The flamelet approach, originally proposed by Spalding \cite{spalding_mixing_1971} and Bilger \cite{bilger_structure_1976}, and later refined by Peters \cite{peters_laminar_1984}, has become one of the standard methods for subgrid-scale combustion modeling. This approach takes advantage of the scale separation that occurs in many turbulent combustion regimes, where chemical timescales are significantly shorter than those associated with turbulent transport, i.e., at high second Damköhler numbers. Under such conditions, the fast chemistry assumption becomes valid, allowing the reaction zone to be treated as a thin structure, termed a flamelet, embedded within the turbulent flow. Turbulent flames can thus be modeled as a collection of laminar flamelets interacting with the resolved flow field. These flamelets, often computed as canonical counterflow diffusion flames \cite{peters_laminar_1984,peters_turbulent_2000}, are solved independently and then coupled to the main CFD solver. This formulation enables the incorporation of detailed reaction mechanisms, including multiple species and reaction steps, without directly solving for all reactive scalars on the resolved scale. As a result, substantial reductions in computational cost can be achieved \cite{nguyen_longitudinal_2018}. 


Flamelet-based modeling has been applied to both premixed and non-premixed combustion. A key element common to both is the generation of a series of flamelet solutions by systematically varying the strain rate imposed at the inflow of the counterflow model. The strain rate (or equivalently, the scalar dissipation rate at the flame front) governs the flame structure determining extinction, ignition, and local heat release through the competition between diffusion and chemical reaction rates.

To couple the subscale flamelet solutions to the resolved-scale simulation, a tracking variable is required to reflect the local flamelet state. Authors approach this coupling differently. In the steady-flamelet model (SLFM) proposed by Cook et al. \cite{cook_laminar_1997} for LES, it is assumed that the flamelet can be fully described by the mixture fraction $Z$ and the scalar dissipation rate at a reference mixture fraction, typically the stoichiometric value $Z_{st}$. In this formulation, only the upper (stable) branch and the non-reacting branch beyond the extinction limit are retained. The scalar dissipation rate serves as the coupling variable linking the resolved and subgrid scales. On the resolved scale, this quantity is either computed from local gradients of $Z$ or modeled using a turbulence-chemistry closure. See Refs. \cite{peters_laminar_1984,peters_turbulent_2000,claramunt_analysis_2006,poinsot_theoretical_2005} for details.

An important limitation of the steady-flamelet model (SLFM) is its inability to accurately capture ignition and extinction phenomena, due to its reliance on a discontinuous formulation that includes only the stable upper branch and the non-reacting branch beyond the extinction limit. To address this limitation, Pierce and Moin \cite{pierce_progress-variable_2004} introduced the Flamelet Progress Variable (FPV) model. This approach enables a more continuous transition between non-reacting (extinguished) and fully reacting (stable) flamelet states by incorporating the unstable middle branch of the S-curve, thereby providing a more realistic representation of ignition and extinction processes. 

In the FPV framework, inclusion of the unstable branch is achieved by mapping all flamelet solutions (stable, unstable, and extinguished) parametrized by scalar dissipation rate to a progress variable $C$. This variable is typically defined either as the maximum temperature or as the sum of selected product species mass fractions, providing a monotonic parameter that reflects the state of chemical progress along the "s-shaped" curve. Coupling to the resolved-scale flow field is then achieved by solving a transport equation for the progress variable on the resolved scale.

Initial formulations of the FPV model were derived under the low-Mach number assumption, rendering them unsuitable for high-speed compressible flows where thermodynamic nonlinearity due to compressibility and viscous dissipation significantly affects the resolved-scale temperature field. To address this deficiency, compressible extensions of the FPV framework have been developed. These extensions reconstruct the resolved temperature via analytical expressions obtained through asymptotic expansions of flamelet-scale thermochemical quantities \cite{pecnik_reynolds-averaged_2012,saghafian_efficient_2015}. Steady compressible flamelet formulations, when combined with the FPV methodology or its derivatives, have emerged as the predominant modeling strategy for diffusion-dominated flames, surpassing classical steady flamelet approaches in both usage and citation frequency, as evidenced by numerous recent studies. See, for example, Refs.  \cite{nguyen_driving_2017,nguyen_impacts_2018,nguyen_longitudinal_2018,nguyen_spontaneous_2019,shadram_neural_2021,shadram_physics-aware_2022,zhan_combustion_2024,pecnik_reynolds-averaged_2012,saghafian_efficient_2015,shan_improved_2021,jiang_species-weighted_2023,coclite_numerical_2015}.

In both compressible and low-Mach-number FPV formulations, the progress-variable transport equation is convective–diffusive, with a source term obtained from the flamelet tables as the sum of the production rates of the species defining $C$. A persistent concern is that this equation is governed primarily by chemistry, with no explicit dependence on the resolved-scale strain rate. Consequently, the evolution of $C$ may not reflect the local strain rate that, in reality, controls the flamelet response.

This study originates from a fundamental question: although the flamelet structure is determined by its applied strain rate, can the resolved-scale progress variable, which lacks direct dependence on strain rate, reliably communicate local mixing conditions to the subgrid flamelet? Ideally, high resolved-scale strain should correspond to a flamelet state characterized by high scalar dissipation rate, and vice versa; however, this coupling may not be guaranteed within the FPV approach.

In a recent study, Walsh et al. \cite{walsh_turbulent_2025} employed a compressible FPV model to simulate an accelerating mixing-layer diffusion flame representative of combustion in turbine-burner environments. Their framework included an additional flamelet library dimension to account for the sensitivity of subgrid-scale chemistry to local pressure variations. Crucially, they identified a structural limitation inherent to progress-variable mappings: once flamelet solutions are projected from scalar dissipation rate to progress variable, the connection between the flamelet-level strain rate and the resolved-scale strain rate magnitudes is lost. As they noted, “Once the subgrid flamelet solutions are mapped to the progress variable, the connection between the flamelet strain rate and resolved-scale strain rate is lost. There is no direct relationship between the resolved-scale mean progress variable value $C$ and the resolved-scale strain rate magnitude.” This observation is critical, as the strain rate represents a key mechanical constraint that governs the flamelet's thermal and chemical state, and therefore, the resolved-scale heat release. 

 Sirignano et al. \cite{sirignano_flamelet_2024} proposed a scaling framework that begins with the turbulence kinetic energy dissipation rate $\epsilon$ as obtained from RANS or LES simulations. Their analysis relates $\epsilon$ to the local turbulent kinetic energy and the viscous dissipation rate at the Kolmogorov scale, enabling an estimation of the smallest eddy turnover times and associated strain rates. Through this approach, the strain rate imposed on the flamelet (representing the mechanical constraint governing its structure) can be determined in a manner consistent with the local turbulence cascade. In this framework, $\epsilon$ becomes the flamelet tracking variable, that is; a given value of $\epsilon$ at a specific spatial and temporal location on the resolved scale is used to infer a subgrid-scale strain rate that reflects the relevant length and time scales of the flamelet. This work was, however, limited to flamelet-level computations; resolved-scale simulations employing this approach have not yet been performed.
 
The objective of the present study is twofold. First, we aim to investigate the implications of the decoupling between the resolved-scale strain rate and the subgrid-scale flamelet strain rate introduced by the use of progress-variable-based flamelet models. Second, we assess the viability of using the turbulent kinetic energy dissipation rate, $\epsilon$, as a tracking variable in flamelet-based modeling. By relating $\epsilon$ to the subgrid-scale strain rate through turbulence scaling arguments, we seek to establish a more physically consistent coupling between resolved and subgrid quantities in steady flamelet formulations. To pursue these objectives, a series of RANS simulations of a two-dimensional, reacting, accelerating mixing layer are performed using multiple combustion models, including a one-step global kinetics (OSK) model, a conventional FPV approach, and a novel $\epsilon$-based flamelet formulation that uses $\epsilon$ as the tracking parameter.

\section{Numerical Framework}

\subsection{Flow and Computational Configuration} \label{sec:domain}
To investigate the main objectives outlined in Sec. \ref{intro} while avoiding the complexity of more elaborate geometries and applications, we consider a diffusion flame in a two-dimensional, steady, transonic reacting mixing layer formed by two separate incoming reactant streams with differing densities and temperatures. We adopt the configuration originally proposed by Zhu et al. \cite{zhu_numerical_2024}, who studied this flow using a RANS solver with an elliptic formulation, and later used by Walsh et al. \cite{walsh_turbulent_2025}, who employed a RANS solver under the boundary-layer approximation (i.e., a parabolic formulation). In both studies, the mixing layer is subject to a strong favorable pressure gradient of 200 atm/m, imposed to mimic the expanding flow field typical of turbine-burner environments. While modeling the turbine burner \cite{sirignano_performance_1999,liu_turbojet_2001} itself is not the aim of the present study, the configuration developed in these works provides a controlled and practical setup for our investigation.

\begin{figure}
\centering
\begin{subfigure}{.55\textwidth}
  \centering
  \includegraphics[width=1.0\textwidth]{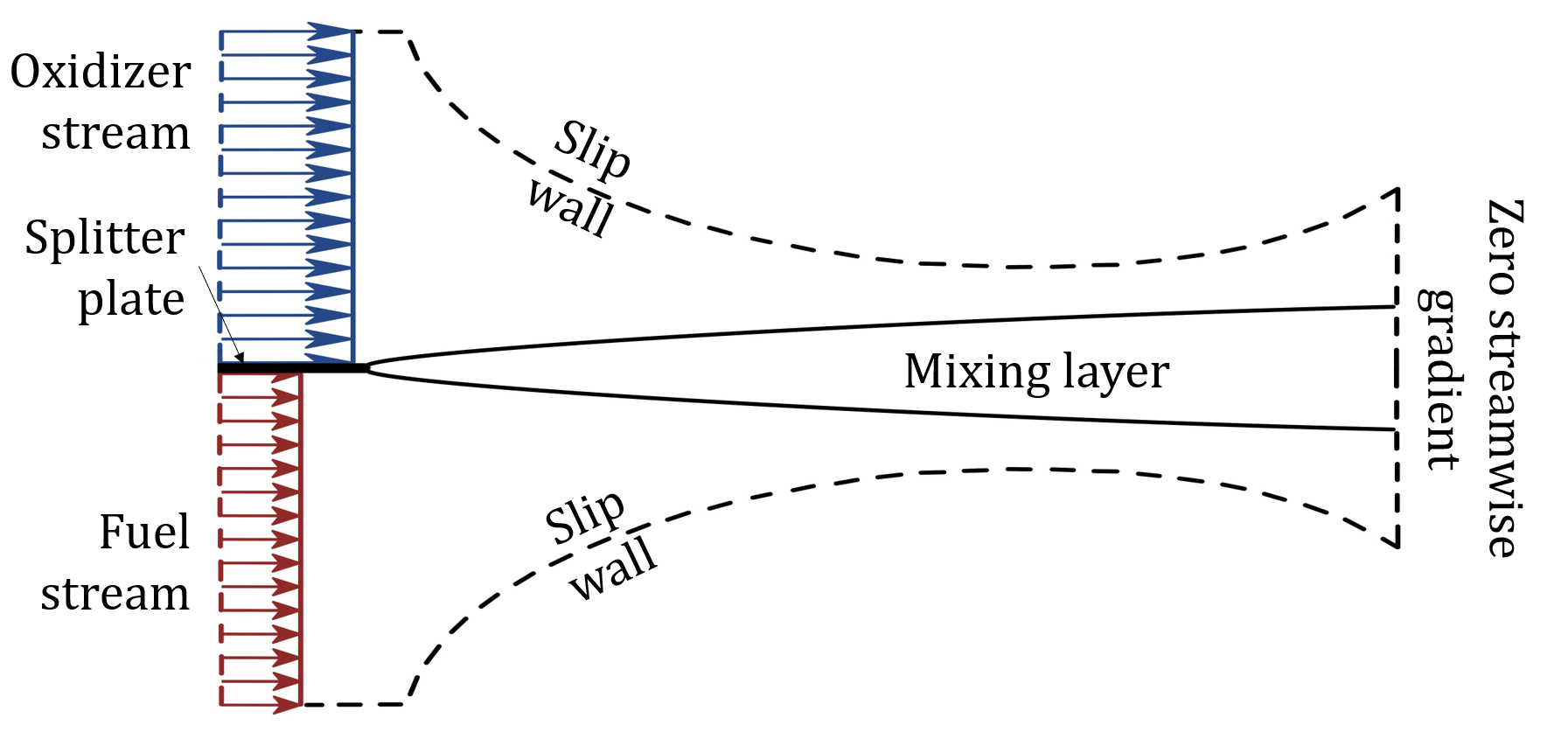}
  \caption{Sketch of the flow configuration.}
  \label{fig:comp_domain}
\end{subfigure}%
\begin{subfigure}{.45\textwidth}
  \centering
  \includegraphics[width=0.97\textwidth]{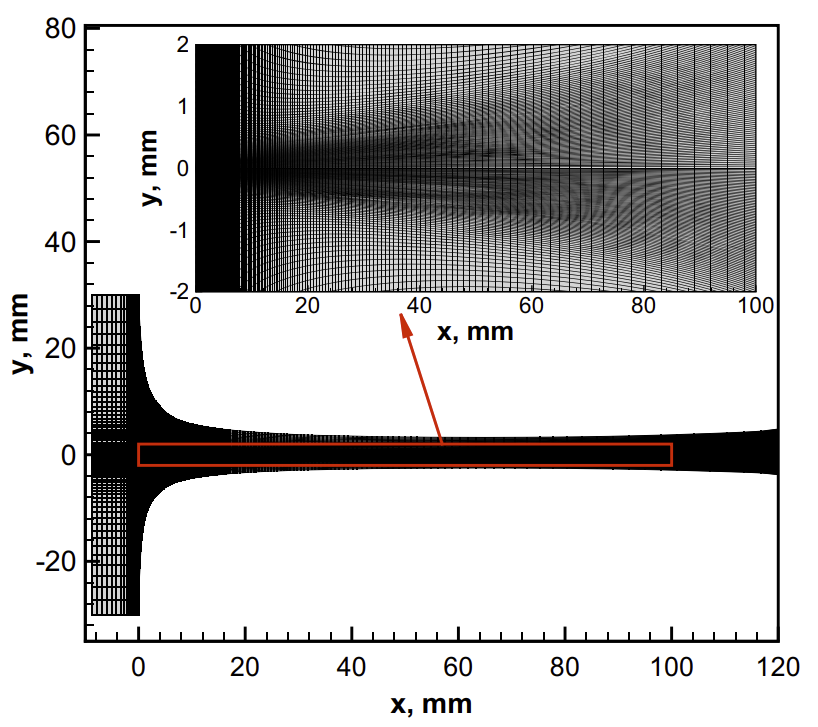}
  \caption{Computational grid.}
  \label{fig:grid}
\end{subfigure}
\caption{Flow configuration and computational grid.}
\label{fig:grid_and_sketch}
\end{figure}

To produce the desired streamwise pressure gradient, Zhu et al. \cite{zhu_numerical_2024} designed a converging–diverging nozzle geometry. At the inlet, a oxidizer stream enters the upper side of the nozzle and comes into contact with vaporized fuel injected from the lower side. The downstream profiles of the upper and lower walls are determined from the isentropic relations for quasi-one-dimensional flow of air and fuel, based on the inlet flow conditions and initial nozzle height. Since this approach assumes inviscid, non-reacting, and unmixed flow, discrepancies arise between the prescribed pressure distribution and the actual pressure field computed in the reacting mixing layer. These discrepancies are corrected iteratively by reshaping the nozzle contours using isentropic relations to match the target pressure distribution.

Figure \ref{fig:comp_domain} illustrates the converging–diverging nozzle geometry used in this configuration. To minimize the influence of inlet boundary conditions on the developing mixing layer, a uniform inlet section is introduced upstream of the converging zone. The upper and lower nozzle walls are nearly symmetric: both converge rapidly downstream of the inlet, flatten in the middle section, and then gradually diverge past the throat located at $x=70$ mm.

The fuel stream consists of pure methane (100\% CH$_4$) at a temperature of 400 K, while the oxidizer stream is composed of an equal mass fraction of oxygen and nitrogen (50\% O$_2$, 50\% N$_2)$ at 1650 K. This oxidizer composition differs from that used by Zhu et al.~\cite{zhu_numerical_2024}, who employed air as the oxidizer. The turbulent intensity and the turbulent-to-molecular viscosity ratio at the inlets are set to 5\% and 10, respectively, for the air stream, and to 10\% and 100 for the fuel stream. The static pressure at the inlets is fixed at 30 bar, and the total pressure is specified to yield inlet velocities of 50 m/s for the fuel and 100 m/s for the air.
Inviscid slip with a zero normal pressure gradient and adiabatic wall boundary conditions are specified on the two side surfaces of the nozzle. All flow quantities are extrapolated at the outlet using a zero streamwise gradient, consistent with the supersonic outflow boundary condition, thereby preventing wave reflections into the computational domain. A symmetric plate segment is included along the centerline between the inlet and the onset of nozzle convergence to suppress premixing between the fuel and oxidizer streams.

The computational grid used in this study was developed by Zhu et al.~\cite{zhu_numerical_2024} and is shown in Fig.~\ref{fig:grid}, along with a zoomed-in view highlighting local details (top). A multiblock structured grid with matched interfaces between neighboring blocks is employed within the nozzle. The vertical grid lines are clustered near the inlet, while the streamwise grid lines are concentrated along the centerline and gradually stretched toward the sidewalls. The height of the first cell adjacent to the centerline is less than 0.05 mm. The grid density was established through a grid-independence study, leading to the selection of a final grid containing 38,272 cells.

\subsection{Governing Equations}
The governing equations required for the compressible, multi-species, reacting RANS computations using all three combustion models are provided hereafter. 
\subsubsection{Reynolds-Averaged Navier-Stokes Equations}
The Reynolds-Averaged Navier-Stokes (RANS) equations for compressible flows with $N$ individual species are expressed by the following transport equations for mass, momentum and energy:
\begin{subequations} \label{eq:navier-stokes}
    \begin{align}
        \frac{\partial\bar{\rho}}{\partial t} + \nabla \cdot (\bar{\rho}\widetilde{\mathbf{V}}) & 
            = 0 \label{eq:continuity} \\
        \frac{\partial(\bar{\rho} \widetilde{\mathbf{V}})}{\partial t} + \nabla \cdot (\bar{\rho}\widetilde{\mathbf{V}} \widetilde{\mathbf{V}}) &
            = -\nabla \bar{p} + \nabla \cdot \boldsymbol{\tau} \label{eq:momentum} \\
        \frac{\partial(\bar{\rho} \widetilde{E})}{\partial t} + \nabla \cdot (\bar{\rho} \widetilde{E}\widetilde{\mathbf{V}}) & = -\nabla \cdot (\bar{p} \widetilde{\mathbf{V}}) + \nabla \cdot (\widetilde{\mathbf{V}} \cdot \boldsymbol{\tau}) - \nabla \cdot \mathbf{q} + \widetilde{\dot Q} \label{eq:energy} 
    \end{align}
\end{subequations}

Here $\widetilde{E}$ is the total sensible energy given by
\begin{equation}
    \widetilde{E} = \tilde{h} - \frac{\bar{p}}{\bar{\rho}} + \frac{1}{2}\widetilde{\mathbf{V}} \cdot \widetilde{\mathbf{V}}
\end{equation}
with
\begin{equation}
    \tilde{h} = \sum_{n=1}^{N}{\widetilde{Y}_n \tilde{h}_n}\quad\text{and} \quad \tilde{h}_n = \int_{T_{\mathrm{ref}}}^{\widetilde{T}}{C_{p,n}(T)\mathrm{d}T} \:,
\end{equation}
where $C_{p,n}$ is given by NASA polynomials \cite{mcbride_coefficients_1993} and $\widetilde{Y}_n$ is the mass fraction of species $n$. Note that with this definition of enthalpy, the energy transport equation takes a non-conservative form, featuring an explicit mean heat release source term $\widetilde{\dot Q}$, which is provided by the combustion model (as discussed in detail below). 

The perfect gas equation of state is assumed,
\begin{equation}
    \bar{p} = \bar{\rho}R\widetilde{T} \:,
\end{equation}
where $R$ is the gas constant of the mixture, computed by the mass-weighted summation of the gas constant of each species $R_n$ with $R_n = R_0/W_n$.

\subsubsection{Species Transport and Flamelet Model Transport Equations}
In this subsection, we summarize the specific transport equations used by each model. 
For the resolved-scale OSK and the $\epsilon$-based flamelet combustion models, species transport is explicitly resolved through $N$ partial density transport equations:
\begin{equation} \label{eq:species}
    \frac{\partial{\bar{\rho} \widetilde{Y}_n}}{\partial t} + \nabla \cdot (\bar{\rho} \widetilde{Y}_n \widetilde{\mathbf{V}}) = -\nabla \cdot \mathbf{j}_n + \widetilde{\dot{\omega}}_n, \ n = 1, \ 2, \ ..., \ N  \:,
\end{equation}
where $\widetilde{\dot{\omega}}_n$ are the species production rates. 

In contrast, the FPV combustion model eliminates the need for these resolved-scale species transport equations. Instead, species mass fractions are obtained directly from a precomputed flamelet library. Coupling between the flamelet solution and the resolved flow field is achieved through the transport of the mean mixture fraction $\widetilde{Z}$ and its mean variance $\widetilde{Z^{''2}}$ . These quantities are governed by their respective transport equations:

\begin{equation}
\frac{\partial \bar{\rho} \widetilde{Z}}{\partial t} + \nabla\cdot(\bar{\rho} \widetilde{Z}\widetilde{\mathbf{V}}) = -\nabla\cdot\mathbf{j}_z
\label{zequation}
\end{equation}

\begin{equation}
\frac{\partial \bar{\rho} \widetilde{Z^{''2}}}{\partial t} + \nabla\cdot(\bar{\rho} \widetilde{Z^{''2}}\widetilde{\mathbf{V}}) = -\nabla\cdot\mathbf{j}_{z^{''2}}+
2\frac{\mu_{T}}{\mathrm{Sc}_{T}}\nabla\widetilde{Z}\cdot\nabla\widetilde{Z}-\bar{\rho}\widetilde{\chi}
\label{varzequation}
\end{equation}
where the final term in the variance equation represents the mean scalar dissipation rate. This quantity is modeled by relating the integral scalar time-scale and the turbulent flow time-scale, yielding an expression proportional to the turbulent dissipation rate and the mixture-fraction variance:
\begin{equation}
    \widetilde{\chi}=C_{\chi}\frac{\epsilon}{k}\widetilde{Z''^2}
\end{equation}
Here, $k$ and $\epsilon$ are the turbulent kinetic energy and turbulent dissipation rate, respectively \cite{peters_turbulent_2000}. $C_{\chi}$ is the proportionality constant between timescales, which is commonly set to $C_{\chi}=2.0$ \cite{janicka_two-variables_1979}. 

Additionally, the FPV approach requires an additional non-conserved scalar transport equation to track the resolved-scale value of the mean progress variable $\widetilde{C}$ 

\begin{equation}
\frac{\partial \bar{\rho} \widetilde{C}}{\partial t} + \nabla\cdot(\bar{\rho} \widetilde{C}\widetilde{\mathbf{V}}) =  -\nabla\cdot\mathbf{j}_c+ \widetilde{\dot{\omega}}_{C} \: ,
\label{cequation}
\end{equation}

\noindent where the source term $\widetilde{\dot{\omega}}_{C}$ of which is provided by the FPV library.

\subsubsection{Transport Properties}
The transport terms in the previous equations are given by 
\begin{subequations}
    \begin{align}
        \boldsymbol{\tau} &= 2(\mu + \mu_T) \left[ \mathbf{S} - \frac{1}{3}(\nabla \cdot \widetilde{\mathbf{V}} ) \mathbf{I} \right], \ \mathbf{S} = \frac{1}{2} \left[\nabla \widetilde{\mathbf{V}} + (\nabla \widetilde{\mathbf{V}})^T \right] \:,\\
        \mathbf{j}_n &= -\left( \frac{\mu}{\mathrm{Sc}_n} + \frac{\mu_T}{\mathrm{Sc}_T} \right) \nabla \widetilde{Y}_n \:,\\
        \mathbf{q} &= -\left( \frac{\mu}{\mathrm{Pr}} + \frac{\mu_T}{\mathrm{Pr}_T} \right) \left( \nabla \tilde{h} - \sum_{n=1}^{N}{\tilde{h}_n \nabla \widetilde{Y}_n} \right) + \sum_{n=1}^{N}{\tilde{h}_n\mathbf{j}_n} \label{eq:heat_flux} \:,\\ 
        \mathbf{j}_z &= -\left(\frac{\mu}{\mathrm{Sc}}+\frac{\mu_{T}}{\mathrm{Sc}_{T}}\right)\nabla\widetilde{Z} \:,\\
        \mathbf{j}_{z^{''2}} &= -\left(\frac{\mu}{\mathrm{Sc}}+\frac{\mu_{T}}{\mathrm{Sc}_{T}}\right)\nabla\widetilde{Z^{''2}} \quad \text{and} \\
        \mathbf{j}_c &= -\left(\frac{\mu}{\mathrm{Sc}}+\frac{\mu_{T}}{\mathrm{Sc}_{T}}\right)\nabla\widetilde{C} \:.
    \end{align}
\end{subequations}
The molecular viscosity $\mu$ is computed by the mass-weighted summation of molecular viscosity of each species given by Lennard-Jones potentials and implemented using Chemkin subroutines. All species Schmidt numbers, $\mathrm{Sc}_n$, are assumed to be 1.0. The turbulent Schmidt number $\mathrm{Sc}_T$, the Prandtl number $\mathrm{Pr}$, and the turbulent Prandtl number $\mathrm{Pr}_T$ are all set as 0.7 in the present study. The last term in Eq. (\ref{eq:heat_flux}) accounts for the energy transport due to mass diffusion of each species with different enthalpy.

\subsubsection{Turbulence Model}
The turbulent viscosity $\mu_T$ is determined by the $k\mbox{-}\omega$ Shear-Stress Transport (SST) model presented by Menter et al. \cite{menter_ten_2003} in 2003:
\begin{subequations} \label{eq:k-omega}
    \begin{align}
        \dfrac{\partial \bar{\rho} k}{\partial t} + \nabla\cdot(\bar{\rho} k \widetilde{\mathbf{V}}) &= P - \beta^*\bar{\rho} k\omega + \nabla \cdot [(\mu+\sigma_k\mu_T) \nabla k] \\
        \dfrac{\partial \bar{\rho} \omega}{\partial t} + \nabla\cdot(\bar{\rho}\omega \widetilde{\mathbf{V}}) &= \dfrac{\gamma\bar{\rho}}{\mu_T}P - \beta\bar{\rho}\omega^2 + \nabla\cdot\left[(\mu+\sigma_\omega\mu_T)\nabla\omega\right] + 2(1-F_1)\dfrac{\bar{\rho}\sigma_{\omega2}}{\omega}\nabla k\cdot \nabla\omega
    \end{align}
\end{subequations}
The production term is
\begin{equation} \label{eq:turbulent_Pk}
    P = \mathrm{min}(\mu_TS^2, 10\beta^*\bar{\rho} k\omega) \:,
\end{equation}
where $S = \sqrt{2\mathbf{S}:\mathbf{S}}$ is the magnitude of the strain-rate tensor $\mathbf{S}$.
The turbulent viscosity is then computed by 
\begin{equation} 
    \mu_T = \dfrac{a_1\bar{\rho} k}{\mathrm{max}\left(a_1 \omega, F_2S\right)} \: .
\end{equation}

The definitions of the blending functions $F_1$ and $F_2$, as well as the model constants can be found in Ref. \cite{menter_ten_2003}. $F_1$ approaches to zero away from the wall, and switches to one inside the boundary layer. $F_2$ is unity for boundary-layer flows and zero for free-shear layers. Both of them are artificially set to zero for the turbulent mixing-layer studied here.

\subsection{Combustion Models}
In this study, we aim to model methane-air combustion governed by the following global reaction
\begin{equation} \label{eq:methane_reaction}
    {\rm CH}_4 + 2 {\rm O}_2 + 7.52 {\rm N}_2 \rightarrow {\rm CO}_2 + 2 {\rm H}_2{\rm O} + 7.52 {\rm N}_2 
\end{equation}
Three different combustion models are employed and detailed below. 

\subsubsection{One-Step Kinetics}
A resolved-scale one-step kinetic combustion model is used as a benchmark case. The chemical reaction rates are determined at the resolved scale using the modified Arrhenius expression:
\begin{equation} \label{eq:Arrhenius_expression}
    {\widetilde{\varepsilon}} = A \widetilde{T}^\beta e^{-E_a/(R_0 \widetilde{T})} \widetilde{C}_{\rm{CH_4}}^a \widetilde{C}_{\rm{O_2}}^b 
\end{equation}
Here, the concentrations are given by $\widetilde{C}_n = \bar{\rho} \widetilde{Y}_n/W_n$. According to Westbrook and Dryer \cite{westbrook_chemical_1984}, $A = 1.3 \times 10^{9} \, \rm{s^{-1}}$, $\beta = 0$, $E_a = 202.506 \, \rm{kJ/mol}$, $a = -0.3$, and $b = 1.3$ for methane/air combustion. Here, species mass fractions for $\mathrm{CH_4}$, $\mathrm{O_2}$, $\mathrm{N_2}$, $\mathrm{CO_2}$ and $\mathrm{H_2O}$ (so $N=5$) are given by the resolved-scale species transport equations (Eqs. \ref{eq:species}). Then, species production rates are determined using stoichiometric coefficients 
\begin{equation}
    \widetilde{\dot\omega}_n = W_n(v_n^{\prime\prime} - v_n^\prime) \widetilde{\varepsilon} \:,
\end{equation}
where $v_n^\prime$ is the stoichiometric coefficient for reactant $n$ in Eq. (\ref{eq:methane_reaction}), and $v_n^{\prime\prime}$ is the stoichiometric coefficient for product $n$.

The mean heat source term (or heat release rate, HRR) $\widetilde{\dot Q}$ appearing on the right-hand side of the energy equation (Eq. (\ref{eq:energy})) is defined using resolved-scale species production rates,
\begin{equation}
    \widetilde{\dot Q} = -\sum_{n=1}^{N}{\widetilde{\dot\omega}_n h^0_{f,n}} \:.
\end{equation}

\subsubsection{Flamelet Progress Variable} \label{sec:fpv}
An extension of the low-Mach-number flamelet progress variable (FPV) approach developed by Walsh et al. \cite{walsh_turbulent_2025} for application to compressible flows is employed in the present work. In this approach, flamelet libraries are generated for the specific reactant compositions and boundary temperatures described in Sec. \ref{sec:domain} by solving the system of steady flamelet equations in a counterflow configuration~\cite{peters_laminar_1984}, namely,
\begin{equation}\label{eq:steady_flamelet}
-\rho \frac{\chi(Z)}{2}\frac{\partial^2 \psi_j}{\partial Z^2}=\dot{\omega}_j, \hspace{1mm}j=1,2,...,M+1
\end{equation}
for varying stoichiometric scalar dissipation rates $\chi_{st}=\chi(Z_{st})$ and background pressures, $p$. Here, $Z$ denotes the mixture fraction and $\psi_j$ represents the reactive scalars, consisting of the mass fractions of the $M$ species considered in the reaction mechanism and temperature. The source terms $\dot{\omega}_j$ correspond to species reaction rates for the mass fraction equations and to the heat release rate in the energy equation. 

The momentum equations do not appear explicitly in the system above because the scalar dissipation rate, $\chi(Z)$, is prescribed in the following canonical form:
\begin{equation}\label{eq:chi_form}
\chi(Z)=\frac{2S^*}{\pi}\exp{(2\mathrm{erfc}^{-1}(2Z)^2)}
\end{equation}
This formulation is widely adopted in the literature \cite{peters_turbulent_2000}, although more comprehensive flamelet formulations that additionally solve the momentum equations have been proposed \cite{sirignano_three-dimensional_2022,sirignano_inward_2022,hellwig_vortex_2025,hellwig_three-dimensional_2025}. In Eq. (\ref{eq:chi_form}), $S^*$ denotes the imposed strain rate at the counterflow inlet. This formulation slightly differs from Peters \cite{peters_turbulent_2000} by a factor of the square root of the ratio of the two densities for the incoming streams, because, we use the strain rate of the incoming fuel stream $S^*$ rather than the oxidizer stream. Notably, $\chi(Z)$ is uniquely parameterized by $S^*$, which in turn is proportional to the stoichiometric scalar dissipation rate $\chi_{st}$.

Solving Eq.~\ref{eq:steady_flamelet} for a range of $\chi_{st}$ and $p$ values yields a steady flamelet library of the form
\begin{equation}\label{eq:psi_chi}
\psi_j=\psi_j(Z,\chi_{st},p)
\end{equation}

\noindent which tabulates the dependence of the reactive scalars on mixture fraction, local strain rate, and background pressure.

These flamelet solutions are computed using the FlameMaster code \cite{pitsch_flamemaster_2022} prior to the resolved-scale computations. The chemical mechanism employed is 13-species, 32-reaction skeletal reduction of Version 1.0 of the Foundational Fuel Chemistry Model (FFCM-1) \cite{smith_foundational_2016,tao_critical_2018}, previously used in FPV computations \cite{zhan_combustion_2024,walsh_turbulent_2025}. The species considered in the skeletal reduction are as follows: $\mathrm{H_2}$, $\mathrm{H}$, $\mathrm{O_2}$, $\mathrm{O}$, $\mathrm{OH}$, $\mathrm{HO_2}$, $\mathrm{H_2O}$, $\mathrm{CH_3}$, $\mathrm{CH_4}$, $\mathrm{CO}$, $\mathrm{CO_2}$, $\mathrm{CH_2O}$ and $\mathrm{N_2}$. Nitrogen is treated as inert, meaning it participates in elementary reactions only as a third body. 

\begin{figure}[]
\centering
\begin{subfigure}{.45\textwidth}
  \centering
  \includegraphics[width=1.0\textwidth]{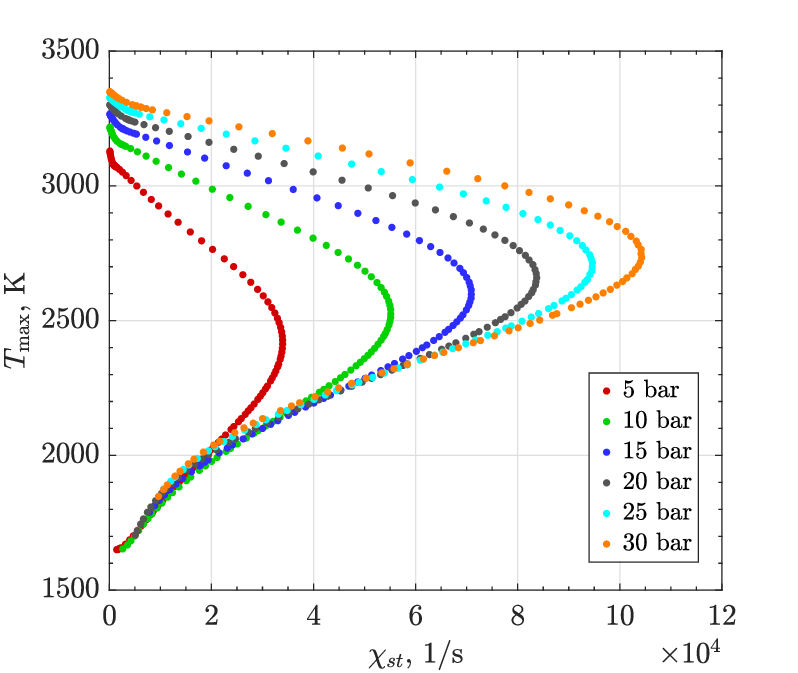}
  \caption{S-shaped curves.}
  \label{fig:s-shaped}
\end{subfigure}%
\begin{subfigure}{.45\textwidth}
  \centering
  \includegraphics[width=0.97\textwidth]{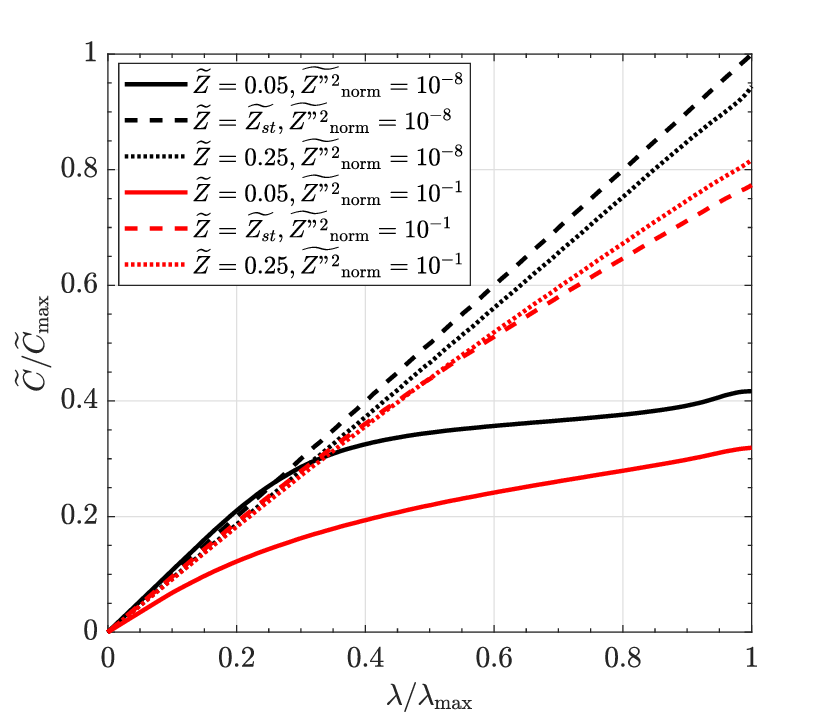}
  \caption{$\mathbf{\widetilde{C}_{\mathrm{tab}}(\boldsymbol{\lambda})=\widetilde{C}(\boldsymbol{\lambda};\widetilde{Z},\widetilde{Z''^2},\bar{p})}$ relations.}
  \label{fig:c_lambda}
\end{subfigure}
\caption{Left: Solutions to the flamelet equations presented as S-shaped curves for various background pressures. Right: $\mathbf{\widetilde{C}_{\mathrm{tab}}(\boldsymbol{\lambda})=\widetilde{C}(\boldsymbol{\lambda};\widetilde{Z},\widetilde{Z''^2},\bar{p})}$ relations for different combinations of $\mathbf{\widetilde{Z}}$ and $\mathbf{\widetilde{Z''^2}}$ at a pressure of 30 bar.}
\label{fig:s-and-pv}
\end{figure}

Figure~\ref{fig:s-shaped} illustrates the steady flamelet solutions in terms of the well-known S-shaped curve, where the ordinate represents the maximum flame temperature (i.e., the peak value of temperature across mixture fraction $Z$) and the abscissa corresponds to the stoichiometric scalar dissipation rate (which is proportional to the imposed inflow strain rate). Flamelet calculations are performed at background pressures of 5, 10, 15, 20, 25, and 30 bar, representing the range of pressures expected at the resolved scale. Each curve exhibits the characteristic bistable structure of non-premixed flames, consisting of an upper stable burning branch and a lower unstable branch, separated by the flammability limit (the point corresponding to the maximum attainable scalar dissipation rate or strain rate) beyond which the flame quenches. As shown in the figure, the flammability limit increases with increasing pressure due to the higher reaction rates.

At this stage, a progress variable $C$ is introduced to provide a monotonic mapping of the multi-branched flamelet solutions onto a single flamelet parameter, $\lambda$. This parameter is defined as the value of the progress variable at a chosen reference mixture fraction, typically the stoichiometric value, such that $\lambda = C(Z_{st})$. The specific definition of $C$ is somewhat ad hoc and depends on the choice of reactants and boundary thermochemical conditions. However, it must guarantee a bijective relationship between  $C$ and $\lambda$ for all combinations of $Z$ and $p$. In other words, given $C(\lambda) = C(\lambda;Z,p)$, the relation must be invertible, allowing the determination of a unique value of $\lambda$ (and corresponding value value of $\chi_{st}$) from a known $C$. This condition guarantees that every combination of $C$, $Z$, and $p$ maps to a unique flamelet solution. Accordingly, Eq. (\ref{eq:psi_chi}) may be re-described as
\begin{equation}\label{eq:psi_lam}
\psi_j = \psi_j(Z,\lambda,p)
\end{equation}

\noindent where one of the $\psi_j$ is the progress variable $C=C(Z,\lambda,p)$. A common formulation of $C$ is a weighted sum of major product species. In the present study, the progress variable is defined as
\begin{equation}\label{eq:c_definition}
C = Y_{\mathrm{H_2O}} + Y_{\mathrm{CO_2}} + Y_{\mathrm{CO}}.
\end{equation}
Although this definition may be considered conventional, more systematic strategies for selecting the progress variable and its weights have been proposed in the literature~\cite{ihme_regularization_2012,najafi-yazdi_systematic_2012,bojko_formulation_2016}. Nonetheless, the findings of the present study are believed to be insensitive to the specific choice of $C$. 

Once the laminar flamelet solutions are mapped to the progress variable, they are convoluted to obtain Favre-averaged first moments of the reactive scalars using the presumed probability density function (PDF) approach,

\begin{equation}\label{eq:lambda_mean}
\widetilde{\psi}_j(\widetilde{Z},\widetilde{Z''^2},\lambda,\bar{p})=\int_0^1 \psi_j(Z,\lambda,\bar{p})\widetilde{P}(Z,\widetilde{Z},\widetilde{Z''^2})dZ
\end{equation}

In this work, a standard $\beta$-PDF is used for the mixture fraction $Z$, and a Dirac $\delta$-PDF is assumed for the flamelet parameter $\lambda$. For further implementation details, the reader is referred to \cite{walsh_turbulent_2025}. Note that the reactive scalars still remain functions of $\lambda$. The final step in the FPV formulation consists of inverting the bijective relation $\widetilde{C}_{\mathrm{tab}}(\lambda)=\widetilde{C}(\lambda;\widetilde{Z},\widetilde{Z''^2},\bar{p})$. These relations are computed by Eq. (\ref{eq:lambda_mean}), and the subscript "tab" is used to distinguish the precomputed tabulated value of the mean progress variable from the value obtained during the resolved-scale computation via its transport equation. This inversion defines a one-to-one mapping between the quantities $\{\widetilde{Z}, \widetilde{Z''^2}, \widetilde{C}_{\mathrm{tab}}, \bar{p}\}$ and a value $\lambda$ and its associated $\chi_{st}$ value. In other words, each combination of $\widetilde{Z}$, $\widetilde{Z''^2}$, $\widetilde{C}_{\mathrm{tab}}$ and $\bar{p}$ uniquely identifies a single steady flamelet solution of Eq. (\ref{eq:steady_flamelet}), i.e., $\widetilde{\psi}_j = \widetilde{\psi}_j(\widetilde{Z},\widetilde{Z''^2},\widetilde{C}_{\mathrm{tab}},\bar{p}).$
Figure \ref{fig:c_lambda} shows the bijective relation $\widetilde{C}_{\mathrm{tab}}(\lambda)=\widetilde{C} (\lambda;\widetilde{Z},\widetilde{Z''^2},\bar{p})$ for various combinations of $\widetilde{Z}$ and $\widetilde{Z''^2}$ at a pressure of 30 bar. The bijectivity of this mapping has ben ensured by the present choice of the progress-variable definition (Eq.(\ref{eq:c_definition})). 

Typically, this inversion is performed offline prior to the CFD simulation to reduce computational cost and the flamelet tables are stored as  $\widetilde{\psi}_j = \widetilde{\psi}_j(\widetilde{Z},\widetilde{Z''^2},\widetilde{C}_{\mathrm{tab}},\bar{p})$. However, in the present work, the inversion is evaluated online during the resolved-scale computations, allowing for the local $\lambda$ values to be stored for analysis. This requires the tabulation of $\widetilde{C}_{\mathrm{tab}}(\lambda)=\widetilde{C} (\widetilde{Z},\widetilde{Z''^2},\lambda,\bar{p})$, which can be stored as an additional reactive scalar within the flamelet libraries, defined as
\begin{equation}\label{eq:mean_tables}
\widetilde{\psi}_j = \widetilde{\psi}_j(\widetilde{Z},\widetilde{Z''^2},\lambda,\bar{p}).
\end{equation}

At runtime, $\widetilde{Z}$, $\widetilde{Z''^2}$, $\widetilde{C}$ and $\bar{p}$ are obtained by solving their respective resolved-scale transport equations. The Favre-averaged reactive scalars are then retrieved by first determining the flamelet parameter $\lambda$, using the resolved-scale $\widetilde{C}$ as the target in the inversion of $\widetilde{C}_{\mathrm{tab}}(\lambda)$. Once $\lambda$ is obtained, the corresponding quantities are evaluated through quadrilinear interpolation of the stored flamelet tables defined by Eq. (\ref{eq:mean_tables}). Consequently, the retrieved $\widetilde{Y}_n$ ($n=1,\ldots,N$) are used to evaluate thermodynamic and transport properties, while the retrieved source terms $\widetilde{\dot{Q}}$ and $\widetilde{\dot{\omega}}_C$ are provided to the resolved-scale energy equation (\ref{eq:energy}) and progress-variable equation (\ref{cequation}), respectively. Note that in the FPV approach, $N=M$. Here, $\widetilde{\dot{\omega}}_C$ denotes the sum of the production rates of the species that define the progress variable, as given in Eq. (\ref{eq:c_definition}).

At this point, it is worth to compare Eq. (\ref{eq:mean_tables}), representing the flamelet solutions parameterized by the progress variable $\widetilde{C}_\mathrm{tab}$, with Eq. (\ref{eq:psi_chi}) where the flamelet solutions are parameterized by the stoichiometric scalar dissipation rate $\chi_{st}$ (i.e., the strain rate). In the latter formulation, the strain rate naturally arises from the system of flamelet equations and serves as the primary mechanism governing the balance between diffusive transport and chemical reaction rates, thereby determining the flame structure. In contrast, the progress-variable formulation replaces this physically grounded parameterization with one based on $\widetilde{C}$ (or more precisely, $\widetilde{C} \rightarrow \lambda$), which is chemically driven through the source term $\widetilde{\dot{\omega}}_C$ and lacks any explicit dependence on the strain rate. For the FPV model to remain physically consistent, the flamelet parameter $\lambda$ would need to exhibit behavior that correlates with the local strain rate. However, as noted by Walsh et al. \cite{walsh_turbulent_2025} and further demonstrated in Sec. \ref{sec:results}, this correspondence does not hold in practice.

\subsubsection{Epsilon-based Flamelet Model}

To address the aforementioned limitation, the present work seeks to establish a more physically consistent coupling between resolved and subgrid quantities in steady flamelet formulations. This is achieved by employing the turbulent kinetic energy dissipation rate, $\epsilon$, obtained from the resolved-scale transport equations, to infer the local strain rate imposed on the flamelet, following the theoretical framework proposed by Sirignano et al.~\cite{sirignano_flamelet_2024}. In their approach, the inflow strain rate to the counterflow configuration, $S^*$, is related to $\epsilon$ through a gradient-based scaling argument:

\begin{eqnarray}\label{eq:sstar-eps}
S^* = \frac{1}{2}\sqrt{ \frac{C_{vd}\ \epsilon}{ \nu[S_1^2 + 1 - S_1]} }
\end{eqnarray}

\noindent where $C_{vd}$ is a dimensionless coefficient accounting for the distribution of viscous dissipation across a range of the smallest turbulent scales. For the counterflow solution to be physically valid, the constraint $C_{vd} < 1$ must be satisfied. While this coefficient remains to be definitively determined from direct numerical simulation (DNS) data, a value of $C_{vd} = 1$ is adopted in the present study, consistent with the assumptions outlined in \cite{sirignano_flamelet_2024}. The parameter $S_1 = 1/2$ corresponds to the axisymmetric counterflow configuration considered here, and $\nu$ is the molecular kinematic viscosity, obtained from the resolved-scale transport properties. $\epsilon$ is evaluated from the turbulence quantities $k$ and $\omega$ provided by their respective resolved-scale transport equations using the standard relation

\begin{equation}
    \epsilon = C_{\mu} k\omega \:,
\end{equation}

\noindent where $C_{\mu} = 0.09$ is a model constant specified by the $k-\epsilon$ turbulence model \cite{chien_predictions_1982}. With this formulation, for any given spatial and temporal location in the resolved-scale domain, $\epsilon$ is used to compute the corresponding flamelet-scale strain rate in accordance with turbulence cascade scaling. This approach reflects the well-established physical principle that scalar gradients intensify as the turbulent length scale decreases, enabling a more consistent coupling between turbulence structure and flamelet dynamics. Given that $S^*$ is known, the solutions of the flamelet system, Eq. (\ref{eq:steady_flamelet}) presented in Sec. \ref{sec:fpv} may be parametrized in terms of $S^*$ as

\begin{eqnarray}
\psi_j = \psi_j(Z,S^*(\epsilon/\nu),p)
\end{eqnarray}

\begin{figure}
    \centering
    \includegraphics[width=0.45\linewidth]{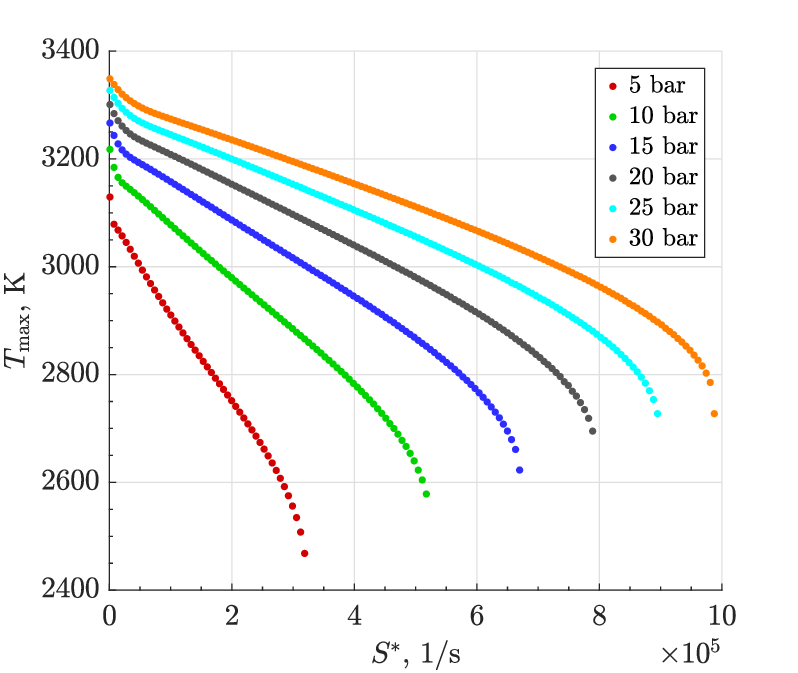}
    \caption{Solutions to the flamelet equations parametrized by $\mathbf{S^*}$.}
    \label{fig:tmax_sstar}
\end{figure}

Figure~\ref{fig:tmax_sstar} presents the same set of flamelet solutions previously shown in Fig.~\ref{fig:s-shaped}, but now parameterized by $S^*$
instead of $\chi_{st}$. Notably, the unstable branch of the classical S-shaped curve is no longer present; only the stable branch extending from small 
$S^*$ to the flammability limit remains. This occurs because $S^*$ alone cannot provide a bijective mapping of both the stable and unstable solutions. Physically, the unstable branch represents an unstable attractor. Small perturbations around these solutions drive the flamelet either toward extinction or back to the stable burning branch. Consequently, such unstable solutions are not expected to persist within an actual turbulent flow. As observed previously, the flammability limit increases with pressure due to the higher chemical rates.

To obtain mean reactive quantities, the Presumed Shape Probability Density Function (PDF) approach used in the FPV framework is applied here as well. The Favre-averaged reactive scalars are obtained through the convolution given in Eq. \ref{eq:lambda_mean}. Here, we employ the standard $\beta$-PDF for the mixture fraction $Z$, while for $\epsilon$ a Dirac $\delta$-PDF is used, reflecting the current absence of a more appropriate statistical description. This assumption may warrant further investigation.

\begin{equation}\label{eq:mean_tables_eps}
\widetilde{\psi}_j = \widetilde{\psi}_j(\widetilde{Z},\widetilde{Z''^2},S^*,\bar{p}).
\end{equation}

During runtime, combinations of the resolved-scale quantities $\widetilde{Z}$, $\widetilde{Z''^2}$, $S^*$ and $\bar{p}$ (where $\widetilde{Z}$ and $\widetilde{Z''^2}$ are obtained by the same resolved-scale transport equations utilized in the FPV approach, i.e., Eqs. (\ref{zequation}) and (\ref{varzequation})), determine the local flamelet state. The corresponding reactive scalars, $\psi_j$, are then retrieved through quadrilinear interpolation of the precomputed tables defined by Eq. (\ref{eq:mean_tables_eps}). If the local strain rate $S^*$ exceeds the flammability limit, the table returns a non-reacting (extinguished) solution.

Unlike the FPV framework, where the reaction progress evolves according to the transport equation for $C$, the present $\epsilon$-based
formulation inherently allows for quenching and re-ignition through its dependence on the dynamically varying local strain rate. Accordingly,
$\epsilon$ should not be interpreted as a progress variable, since its magnitude may increase or decrease in both space and time. Rather, it serves as a coupling variable, linking the resolved-scale turbulence characteristics to the subgrid flamelet dynamics in a physically consistent manner.

However, this quenching mechanism introduces a degree of inconsistency when the species mass fractions are obtained directly from the precomputed flamelet tables. Once the flammability limit is exceeded, the tables return zero product mass fractions, implying complete extinction. Consequently, the non-zero products formed upstream of a locally quenched region are not transported downstream, due to the lack of a resolved-scale transport mechanism. In the FPV formulation, this artifact does not arise because the product species are indirectly advected through the resolved-scale progress variable transport equation. To remedy this issue in the $\epsilon$-based approach, the resolved-scale partial density transport equations (Eqs. (\ref{eq:species})) are retained, while the chemical source terms, $\widetilde{\dot{\omega}}_n$, are retrieved from the flamelet tables defined by Eq. (\ref{eq:mean_tables_eps}). When the local strain rate $S^*$ exceeds the flammability limit, the tables return $\widetilde{\dot{\omega}}_n = 0$ for all species, corresponding to a non-reacting state. This formulation preserves the transport of product species generated upstream, thereby allowing them to convect and diffuse into adjacent regions, even under local quenching conditions.

Introducing resolved-scale species transport equations increases computational cost relative to the FPV formulation, owing to the larger number of scalar equations and the timestep restrictions imposed by stiff chemical source terms. To mitigate this, only a reduced subset of major species is explicitly transported on the resolved scale. Let the detailed reaction mechanism contain $M$ species. We solve transport equations for $P+1 = N$ species, with $N < M$, where the subset of $P$ species corresponds to the dominant contributors to the mixture composition. Defining  
\begin{equation}\label{eq:gamma}
\gamma = \sum_{n=1}^{P} \widetilde{Y}_n,
\end{equation}
the remaining fraction is represented by a composite lumped species,  
\begin{equation}\label{eq:prod}
\widetilde{Y}_{N} = 1 - \gamma,
\end{equation}
with a source term ensuring mass conservation,  
\begin{equation}
\widetilde{\dot{\omega}}_{N} = -\sum_{n=1}^{P} \widetilde{\dot{\omega}}_n \:,
\end{equation}

\noindent so that $\sum_{n=1}^{N} \widetilde{\dot{\omega}}_n = 0$. In the present study, the mechanism comprises \( M = 13 \) species. A reduced set of \( P = 6 \) major species (\(\mathrm{O_2}\), \(\mathrm{H_2O}\), \(\mathrm{CH_4}\), \(\mathrm{CO}\), \(\mathrm{CO_2}\), and \(\mathrm{N_2}\)) is explicitly tracked, while the remaining minor radicals (\(\mathrm{H_2}\), \(\mathrm{H}\), \(\mathrm{O}\), \(\mathrm{OH}\), \(\mathrm{HO_2}\), \(\mathrm{CH_3}\), \(\mathrm{CH_2O}\)) are lumped into $\widetilde{Y}_{N=7}$. This choice captures at least 95\% of the total mass fraction (\(\gamma \ge 0.95\)) across the entire flamelet solution space. In this way, the computational cost of tracking the full set of $M$ species on the resolved scale is avoided, while chemical source terms are still evaluated using subgrid detailed finite-rate kinetics. Furthermore, the timestep constraint is alleviated since the major species exhibit less stiff source terms than the short-lived radicals. 

Finally, the resolved-scale heat release rate, $\widetilde{\dot{Q}}$, is obtained from the flamelet model by interpolating the pretabulated flamelet libraries defined by Eq.~(\ref{eq:mean_tables_eps}).

Table~\ref{tab:summary} summarizes the combustion model formulations and their associated transport equations. Flamelet-based models offer a clear advantage in representing detailed chemical kinetics without a proportional increase in the number of transport equations. For instance, if the same 13-species mechanism used in the flamelet libraries were solved directly on the resolved scale, the system would require 20 transport equations, compared to 10–14 for the flamelet approaches.
The benefit becomes even greater for larger mechanisms. However, the proposed $\epsilon$-based formulation introduces additional cost relative to the FPV model. In FPV, the number of transport equations ($\widetilde{Z}$, $\widetilde{Z''^2}$ and $\widetilde{C}$) is independent of mechanism complexity.
In contrast, the $\epsilon$-based model requires solving a subset of resolved-scale species equations, whose number depends on the mechanism size to maintain $\gamma$ (Eq. (\ref{eq:gamma})) close to unity. For example, Walsh et al. \cite{walsh_flamelet_2026} applied the $\epsilon$-based model to simulate JP-5/air combustion in a turbine passage using a 118-species mechanism and tracked 14 species on the resolved scale. Although this still represents a substantial reduction, the $\epsilon$-based model remains more expensive than FPV. In the present 2-D RANS cases, this added cost is not significant, but its scaling for 3-D LES computations remains to be quantified and may warrant a dedicated cost study.

\setcounter{table}{0}
\begin{table}[h]
\caption{Summary of combustion model formulations and associated transport equations.}
\label{tab:summary}
\begin{tabular}{cccccc}
\cline{2-6}
\multicolumn{1}{c|}{} & \multicolumn{1}{c|}{\textbf{\begin{tabular}[c]{@{}c@{}}Resolved-scale\\  species, $\mathbf{N}$\end{tabular}}} & \multicolumn{1}{c|}{\textbf{\begin{tabular}[c]{@{}c@{}}Resolved-scale \\ species transport\end{tabular}}} & \multicolumn{1}{c|}{\textbf{\begin{tabular}[c]{@{}c@{}}Evaluation of \\ chemical \\ source terms\end{tabular}}} & \multicolumn{1}{c|}{\textbf{\begin{tabular}[c]{@{}c@{}}Reactive scalars \\ $\mathbf{\boldsymbol{\psi}_j}$ brought \\ to the\\  resolved scale\end{tabular}}} & \multicolumn{1}{c|}{\textbf{\begin{tabular}[c]{@{}c@{}}Total resolved\\ -scale\\  transport \\ equations\end{tabular}}} \\ \hline
\multicolumn{1}{|c|}{\textbf{OSK}} & \multicolumn{1}{c|}{5} & \multicolumn{1}{c|}{\begin{tabular}[c]{@{}c@{}}Explicitly, through \\ resolved-scale\\  species transport \\ equations\end{tabular}} & \multicolumn{1}{c|}{\begin{tabular}[c]{@{}c@{}}From a single \\ global reaction\\  based on\\  resolved-scale \\ species concentrations \\ (resolved-scale \\ chemistry)\end{tabular}} & \multicolumn{1}{c|}{N/A} & \multicolumn{1}{c|}{\begin{tabular}[c]{@{}c@{}}12 (5 RANS, \\ 2 $k-\omega$ \\ and 5 species)\end{tabular}} \\ \hline
\multicolumn{1}{|c|}{\textbf{FPV}} & \multicolumn{1}{c|}{$13$*} & \multicolumn{1}{c|}{\begin{tabular}[c]{@{}c@{}}Implicitly, through \\ the resolved-scale \\ $\widetilde{Z}$ and $\widetilde{C}$ \\ equations\end{tabular}} & \multicolumn{1}{c|}{\begin{tabular}[c]{@{}c@{}}From 13 species \\ and 32 reactions \\ solved in the \\ flamelet equations \\ (sub-grid chemistry)\end{tabular}} & \multicolumn{1}{c|}{\begin{tabular}[c]{@{}c@{}}$\widetilde{Y}_{n}$, $\widetilde{C}_{\mathrm{tab}}$***\\ and $\widetilde{\dot{Q}}$\end{tabular}} & \multicolumn{1}{c|}{\begin{tabular}[c]{@{}c@{}}10 (5 RANS, \\ 2 $k-\omega$, \\ $\widetilde{Z}$, $\widetilde{Z''^2}$ \\ and $\widetilde{C}$)\end{tabular}} \\ \hline
\multicolumn{1}{|c|}{\textbf{$\boldsymbol{\epsilon}$-based}} & \multicolumn{1}{c|}{$7$**} & \multicolumn{1}{c|}{\begin{tabular}[c]{@{}c@{}}Explicitly, through \\ resolved-scale \\ species transport \\ equations\end{tabular}} & \multicolumn{1}{c|}{\begin{tabular}[c]{@{}c@{}}From 13 species \\ and 32 reactions \\ solved in the \\ flamelet equations \\ (sub-grid chemistry)\end{tabular}} & \multicolumn{1}{c|}{\begin{tabular}[c]{@{}c@{}}$\widetilde{\dot{\omega}}_{n}$ and $\widetilde{\dot{Q}}$\end{tabular}} & \multicolumn{1}{c|}{\begin{tabular}[c]{@{}c@{}}14 (5 RANS, \\ 2 $k-\omega$, \\ $\widetilde{Z}$, $\widetilde{Z''^2}$\\  and 7 species)\end{tabular}} \\ \hline
\multicolumn{6}{l}{\begin{tabular}[c]{@{}l@{}}* Corresponding to the number of species, $M$, considered in the reaction mechanism used to solve the flamelet equations. \\ ** Corresponding to the subset of the $M$ species that are tracked on the resolved scale. \\ *** $\widetilde{C}_{\mathrm{tab}}$ is the tabular progress variable used for the determination of $\lambda$.\end{tabular}}
\end{tabular}
\end{table}

\subsection{Numerical Solver and Grid}
An in-house three-dimensional code for simulating steady and unsteady transonic flows for single species within turbomachinery blade rows has been developed, validated, and applied by Refs. \cite{zhu_numerical_2017,zhu_flow_2018,zhu_influence_2018,liu_computational_2025}. The code solves the Navier–Stokes equations together with various turbulence models by using the second-order cell-centered finite-volume method based on a multiblock structured grid. The central schemes with artificial viscosity, flux difference splitting schemes, and advection upstream splitting methods with various options to reconstruct the left and right states have been developed and implemented in  the code. 
Recently, it has been extended to include solving transport equations for multiple species with varying specific heat capacities and appropriate chemistry models, and verified and validated by the two-dimensional steady transonic reacting flows in a mixing layer and a turbine cascade \cite{zhu_numerical_2024}, three-dimensional reacting flow in a turbine stage \cite{zhu_large-eddy_2025} and the three-dimensional unsteady reacting flow in a rocket engine \cite{zhu_simulation_2025}. The convective and viscous fluxes are discretized by the Jameson-Schmidt-Turkel scheme \cite{jameson_numerical_1981} and the second-order central scheme, respectively. The local time-stepping method is introduced to accelerate the convergence to a steady state. Thus, the time $t$ in the governing equations is interpreted as a pseudo-time, and a large enough pseudo-time step determined by the local flowfield can be used in each grid cell since time accuracy is not required for steady-state solutions. An operator-splitting scheme is used to treat the stiff chemical source terms in Eqs. (\ref{eq:energy}) and (\ref{eq:species}) (for details see \cite{zhu_numerical_2024}). Parallel techniques based on the message passing interface (MPI) are adopted to further accelerate the computation by distributing grid blocks among CPU processors.

\begin{figure}[!h]
\centering
\begin{subfigure}{.45\textwidth}
  \centering
  \includegraphics[width=1.0\textwidth]{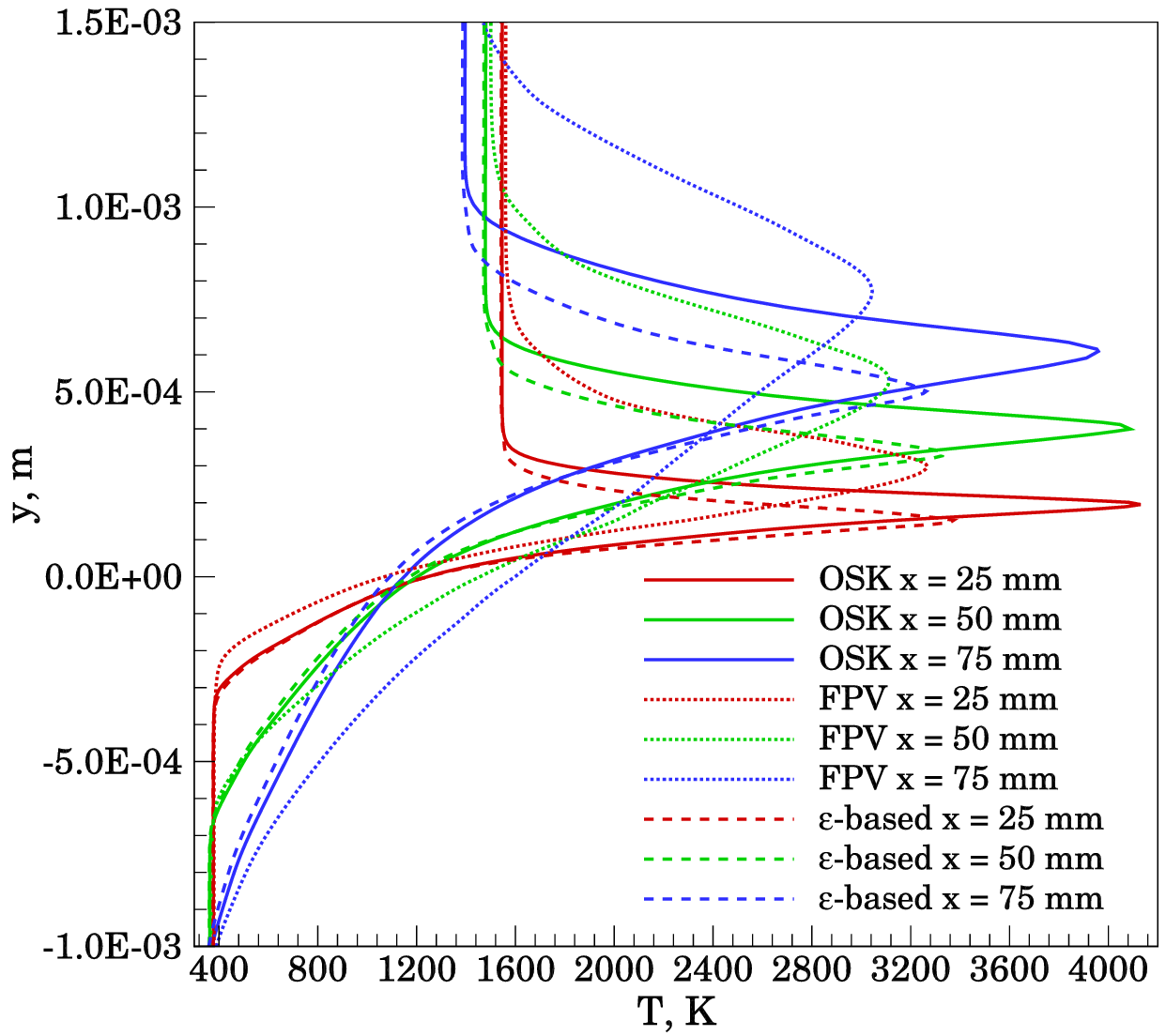}
  \caption{Temperature.}
  \label{fig:temp_lines}
\end{subfigure}%
\begin{subfigure}{.45\textwidth}
  \centering
  \includegraphics[width=1.0\textwidth]{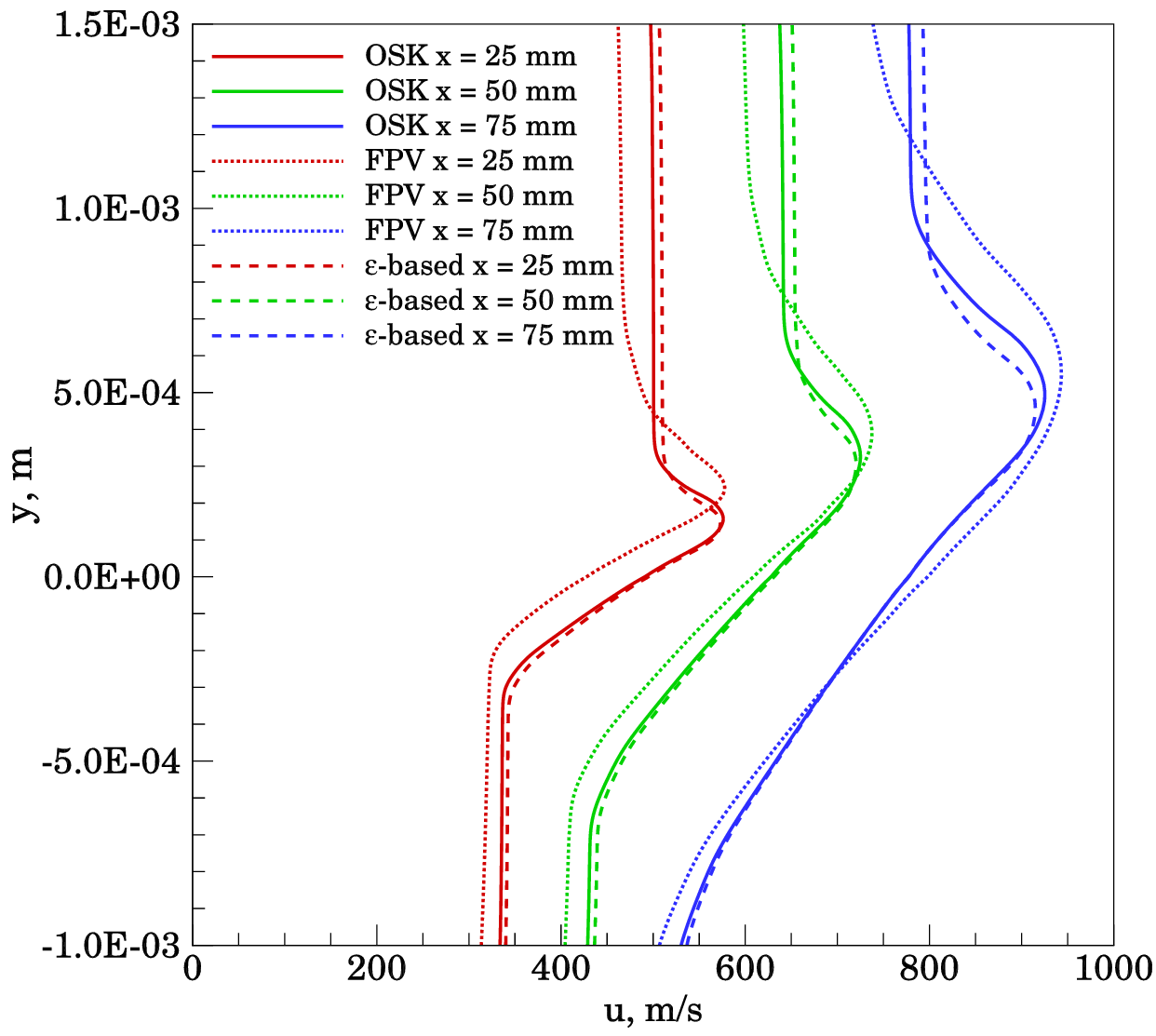}
  \caption{Streamwise velocity.}
  \label{fig:u_lines}
\end{subfigure}
\caption{Profiles of temperature and velocity at three different streamwise locations.}
\label{fig:mixing_layers}
\end{figure}

\section{Computational Results and Discussions}\label{sec:results}
This section presents the results of the reacting flow within the nozzle, obtained using the OSK, FPV and $\epsilon$-based combustion models. At the trailing edge of the splitter plate (x = 3 mm), the hot oxidizer stream above begins to mix with the colder fuel vapor stream below, forming thermal and velocity mixing layers that extend downstream. Ignition occurs where the two streams merge, establishing a diffusion flame within the mixing layer, with the reaction zone located near the stoichiometric mixture composition. Figure \ref{fig:mixing_layers} presents the temperature and streamwise velocity profiles at three downstream positions. The transverse direction has been magnified to highlight the layer development. As the flow progresses, the mixing layer thickens and the freestream velocity increases due to flow acceleration under the large favorable pressure gradient. The air-side mixing layer is noticeably thicker than the fuel-side layer, which can be attributed to the lower density and higher molecular viscosity of the hot oxidizer stream, resulting in a smaller local Reynolds number despite its higher velocity. The thermal and velocity layer thicknesses remain approximately similar, consistent with the use of identical turbulent Schmidt and Prandtl numbers in the simulation.

 Both flamelet-based models predict substantially lower flame temperatures, by approximately 800 K compared to the OSK model. This difference is expected, since flamelet formulations include detailed finite-rate chemistry, which accounts for dissociation losses and other non-equilibrium effects that reduce the adiabatic flame temperature. A similar trend was reported by Walsh et al. \cite{walsh_turbulent_2025}. The FPV results also exhibit significantly thicker thermal and velocity mixing layers than the other two models. This behavior arises because the FPV model predicts a broader reaction zone, where the width of the source terms is governed by the mixture fraction distribution. In contrast, the $\epsilon$-based flamelet model produces mixing-layer thicknesses comparable to those obtained with the OSK formulation, with reaction-zone widths of similar magnitude. A more detailed comparison of these features is presented in the following discussion.

\begin{figure}
    \centering
    \includegraphics[width=0.8\linewidth]{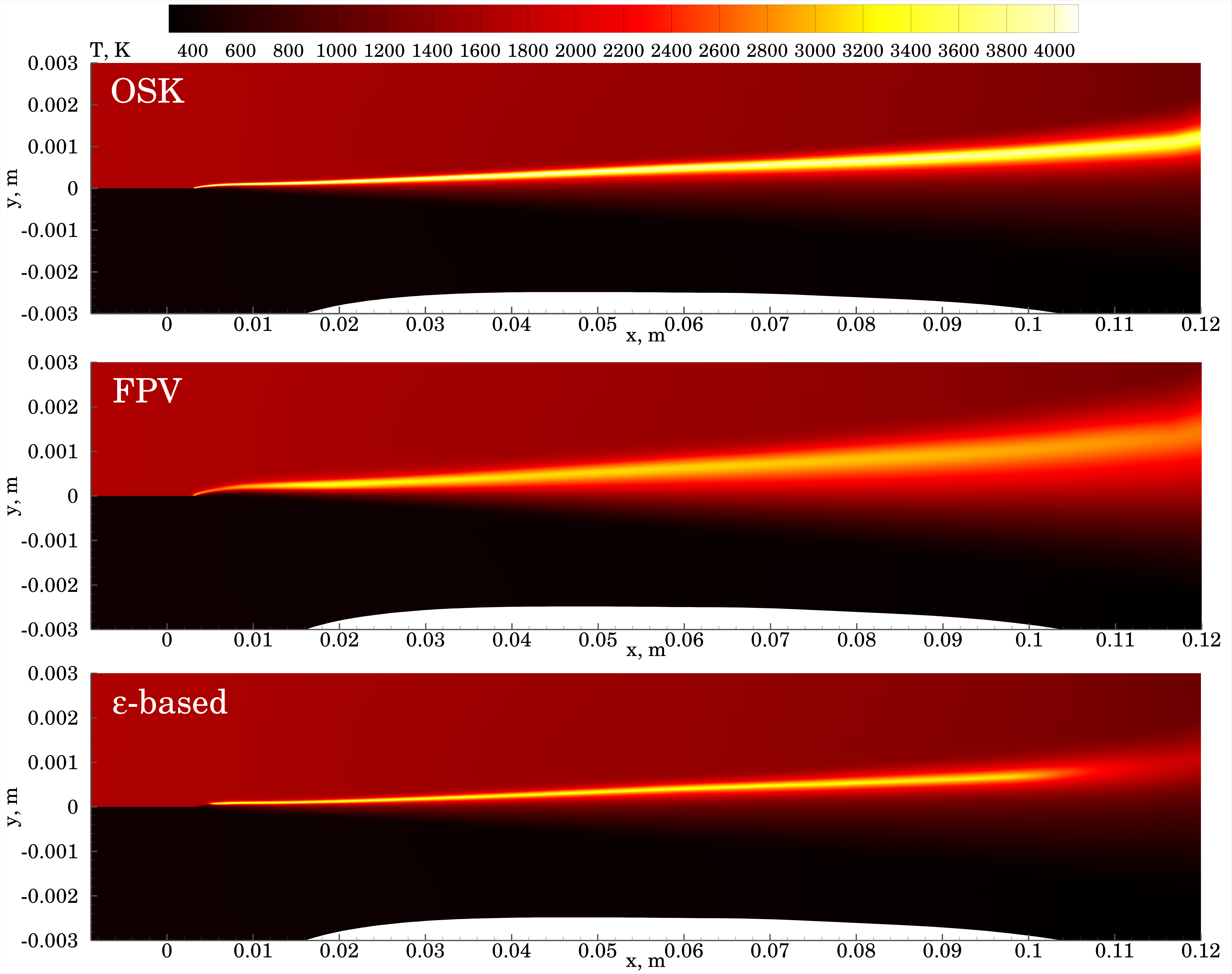}
    \caption{Temperature contours for OSK (top), FPV (center) end $\boldsymbol{\epsilon}$-based (bottom) combustion models.}
    \label{fig:tempcont}
\end{figure}

Figure~\ref{fig:tempcont} presents the temperature fields predicted by the three combustion models. The transverse direction has been magnified to highlight the flame structure in greater detail. Overall, the three models exhibit qualitatively similar behavior, aside from the lower peak temperatures predicted by the flamelet models, as discussed previously. A key difference among the models lies in the flame standoff distance. The OSK model predicts zero standoff, with ignition occurring immediately at the trailing edge of the splitter plate. The FPV model, while also igniting almost immediately, shows a short region of reduced heat release before reaching peak temperatures at approximately x = 10 mm. The $\epsilon$-based flamelet model, in contrast, predicts a distinct flame standoff of about 2 mm.

In the latter case, the flame standoff arises because the local flamelet strain rate, $S^*$, is large in the vicinity of the splitter plate, where strong shear develops due to the no-slip boundary condition of the splitter plate and the large velocity gradients between the two streams. Immediately downstream of the plate, the strain rate exceeds the flammability limit of the flamelet, leading to local quenching. By contrast, although the FPV model also predicts a weak reaction zone downstream of the splitter plate, its behavior depends on the growth rate of the progress variable, which is known to be sensitive to the chosen definition of $C$ (see Eq. (\ref{eq:c_definition})). Different formulations of $C$ produce different source terms $\widetilde{\dot{\omega}}_C$, and consequently yield varying resolved-scale reaction rates and flame standoff distances.

Additionally, the $\epsilon$-based model exhibits a pronounced temperature drop near x = 100 mm, where the static pressure is approximately 10 bar. This drop corresponds to local flamelet quenching, which arises from the decrease in the flammability limit with pressure. Both this downstream quenching and the previously observed flame standoff reflect a direct response of the flamelet to the resolved-scale strain rate, as intended in the $\epsilon$-based formulation. Such behavior cannot be captured by the OSK model, which lacks a mechanism to do so, and is only partially reproduced by the FPV model, as will be discussed shortly.

\subsection{On the Coupling of the Flamelet Models}
We seek to establish a more physically consistent coupling between the resolved and subgrid quantities in steady flamelet formulations; specifically, a model in which the flamelet responds to the resolved-scale strain rate.

\begin{figure}
    \centering
    \includegraphics[width=1.0\linewidth]{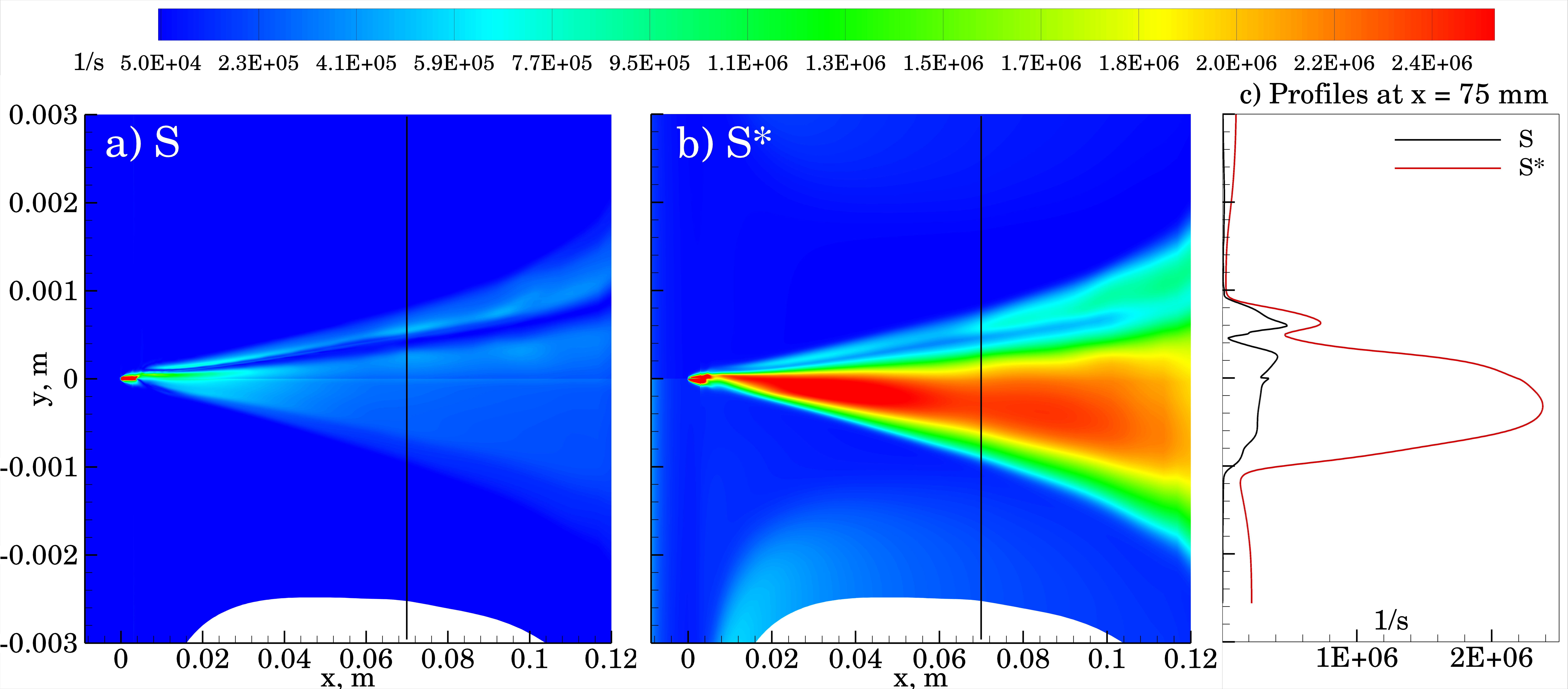}
    \caption{Left (a): contours of resolved-scale strain rate magnitude, $\mathbf{S}$. Center (b): contours of the flamelet inflow strain rate $\boldsymbol{S^*}$. Right (c): profiles of $\mathbf{S}$ and $\mathbf{S^*}$ at the streamwise location of x = 75 mm (represented by the black vertical lines in (a) and (b)).}
    \label{fig:sr_combined}
\end{figure}

Figure~\ref{fig:sr_combined} (a) presents contours of the strain-rate magnitude, $S$, for the $\epsilon$-based simulation. The OSK and FPV results exhibit similar patterns. The splitter plate extends from the inlet to x = 3 mm, but the no-slip boundary condition is applied only between x = 0 mm and x = 3mm. Within this region, the strong velocity gradients induced by the wall generate the highest strain rates, which appear as the red zone in the contour plot. Downstream of the splitter plate, two high-strain regions develop on either side of the flame, separated by a low-strain trough located at the flame centerline. These regions of elevated strain correspond to the strong velocity gradients that form on both sides of the reaction zone, while the local minimum coincides with the flame core, where the velocity derivatives approach zero and strain-rate production is minimized. A profile of $S$ evaluated at x = 75 mm is shown in Fig. \ref{fig:sr_combined} (c).

Figure~\ref{fig:sr_combined} (b) shows the corresponding contours of the flamelet strain rate, $S^*$, computed using the gradient-based scaling argument with $\epsilon$ according to Eq.~(\ref{eq:sstar-eps}). Its profile at x = 75 mm is shown in Fig. \ref{fig:sr_combined} (c). A comparison with the resolved-scale strain rate $S$ reveals that $S^*$ exhibits a similar spatial structure, featuring two high-strain regions flanking the flame and a trough at the reaction zone. This indicates that the subgrid flamelet dynamics respond consistently to the resolved-scale strain field. The magnitudes of $S^*$ are systematically higher than those of $S$, reflecting the increase in strain intensity at smaller turbulent scales, consistent with the energy cascade behavior of turbulence. This behavior is made possible by employing $\epsilon$ as the flamelet tracking variable, which establishes a direct link between the resolved-scale turbulence dissipation and the subgrid flamelet strain rate.

\begin{figure}
    \centering
    \includegraphics[width=0.5\linewidth]{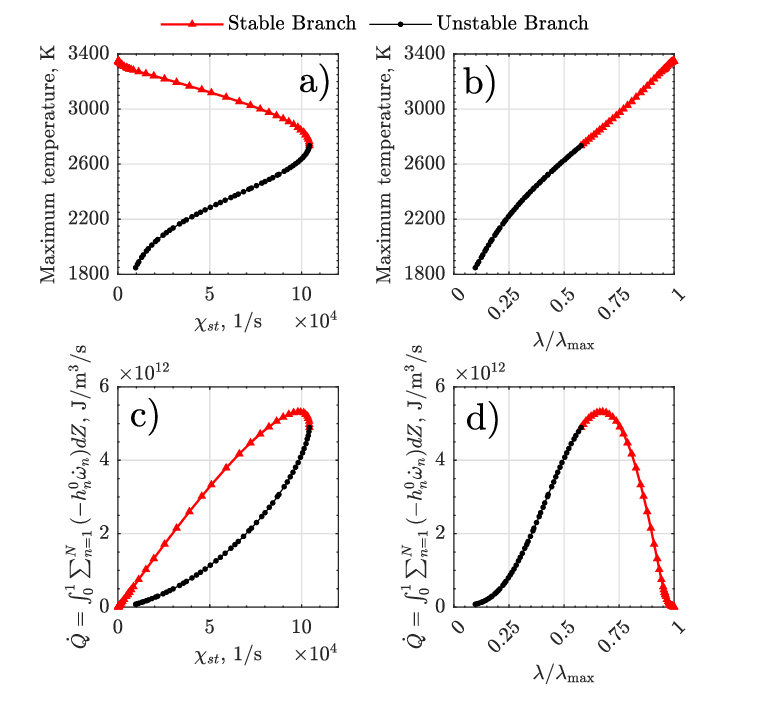}
    \caption{Mapping of the maximum temperature and the heat release rate to the normalized flamelet parameter $\mathbf{\boldsymbol{\lambda}/\boldsymbol{\lambda}_{\mathrm{max}}}$ for a background pressure of 30 bar.}
    \label{fig:qmapping}
\end{figure}

We now examine the coupling between the resolved and subgrid scales in the FPV model. In this formulation, the flamelet strain rate is replaced by the flamelet parameter $\lambda$ when the solutions are mapped from strain-rate space to progress-variable space. The top panels of Fig.~\ref{fig:qmapping} illustrate this transformation, showing the flamelet solutions parameterized by $\chi_{st}$ (which depends on the flamelet strain rate) in (a), and their corresponding representations parameterized by the normalized flamelet parameter $\lambda/\lambda_{\mathrm{max}}$ in (b). The bottom panels show the corresponding mapping of the integrated burning rate, obtained by integrating the local heat-release rate $\dot{Q}$ over the mixture fraction coordinate $Z$: the distribution in terms of $\chi_{st}$ is shown in (c), and the same quantity expressed in terms of $\lambda/\lambda_{\mathrm{max}}$ is shown in (d).

\begin{figure}
\centering
\begin{subfigure}{.5\textwidth}
  \centering
  \includegraphics[width=0.96\textwidth]{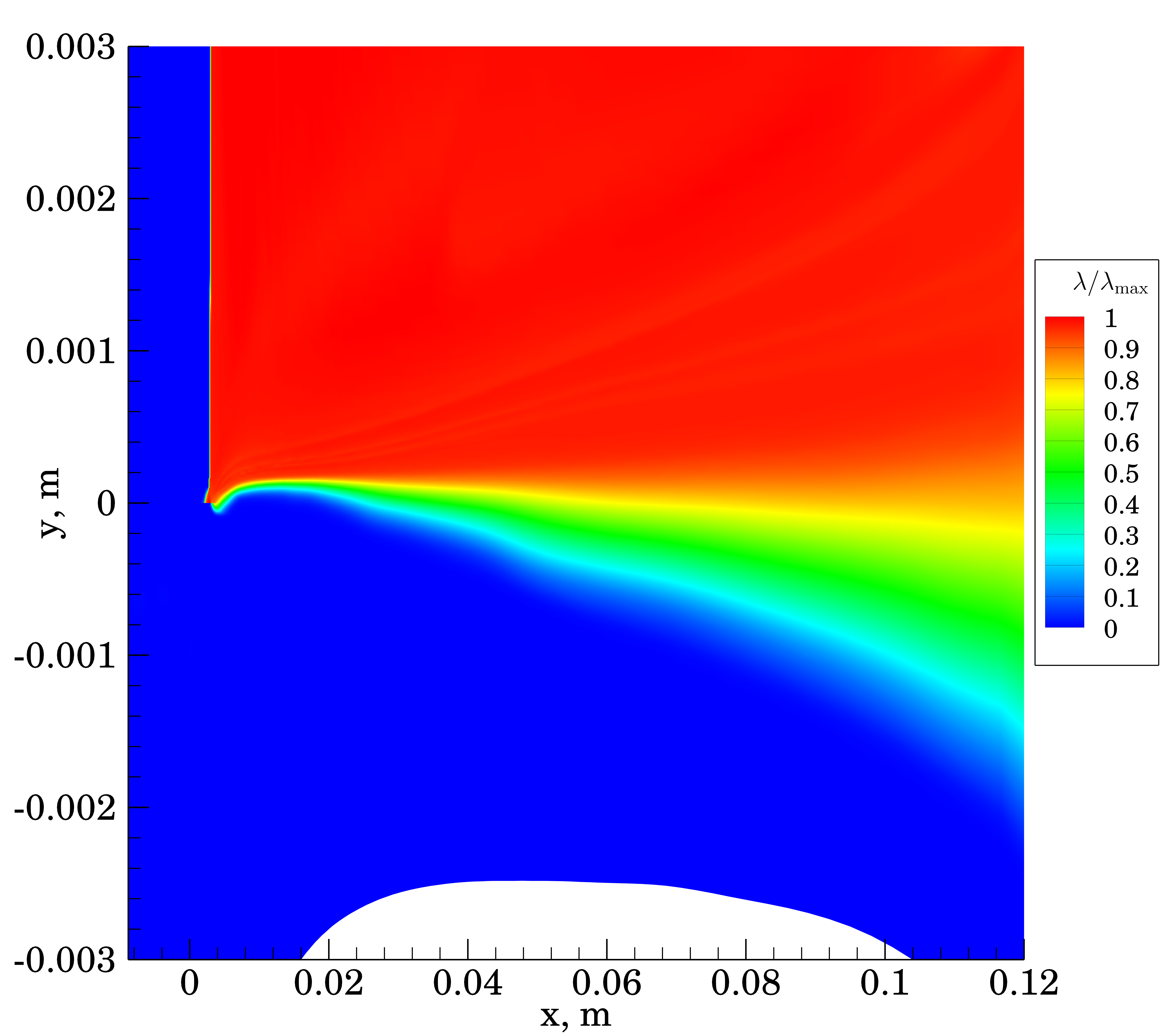}
  \caption{Contour of the normalized flamelet parameter $\boldsymbol{\lambda/\lambda_{\mathrm{max}}}$.}
  \label{fig:fpv_lambda}
\end{subfigure}%
\begin{subfigure}{.5\textwidth}
  \centering
  \includegraphics[width=0.95\textwidth]{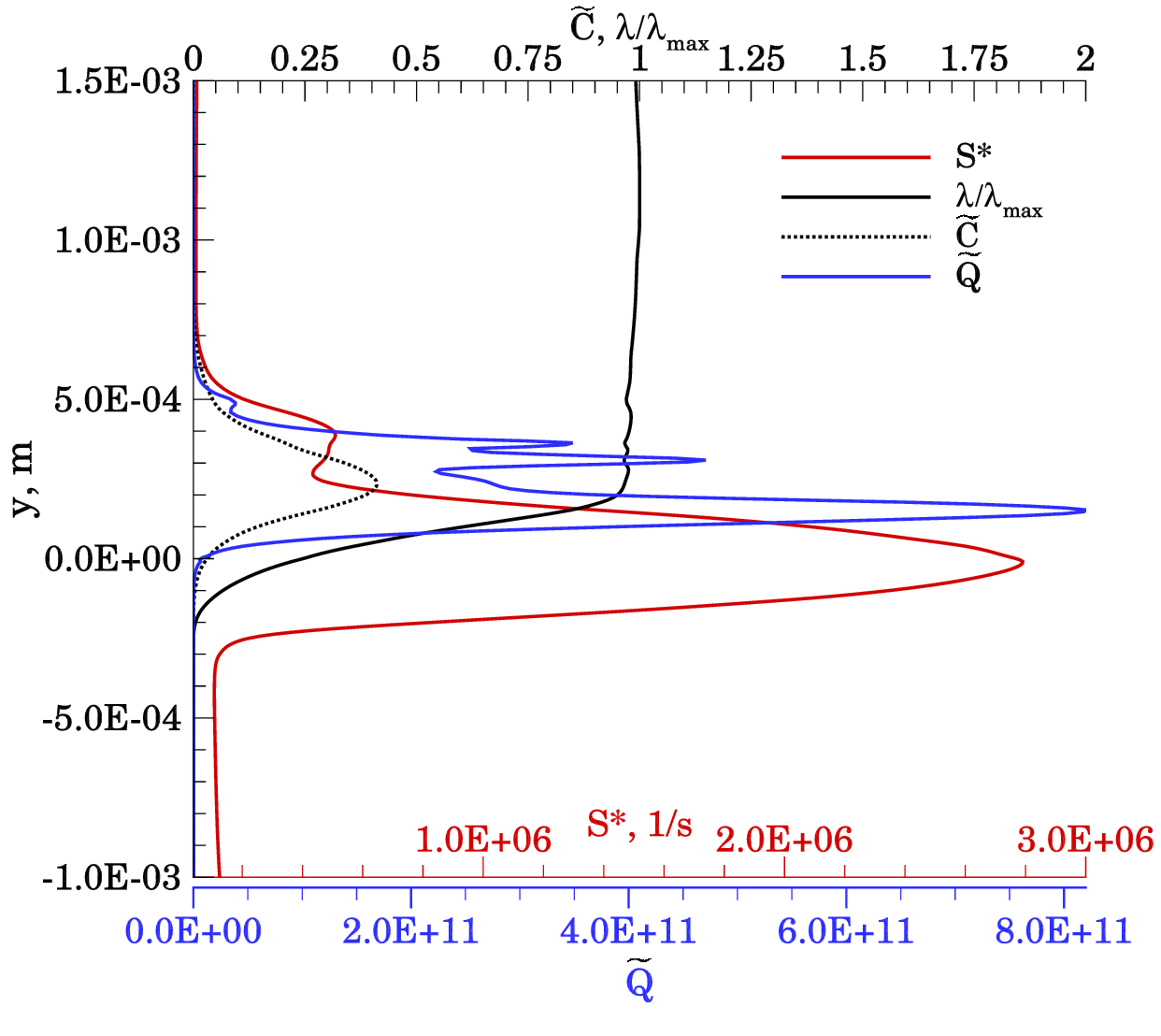}
  \caption{$\boldsymbol{S^*}$, $\boldsymbol{\lambda/\lambda_{\mathrm{max}}}$, $\boldsymbol{\widetilde{C}}$ and $\boldsymbol{\widetilde{\dot{Q}}}$ at x = 25 mm.}
  \label{fig:fpv_lines}
\end{subfigure}
\caption{Quantities of interest for the FPV results.}
\label{fig:fpv_quantites}
\end{figure}

Although the local heat release rate communicated to the resolved scale originates from a single value of $Z$ at each grid point, the integrated heat-release rate provides the relative burning strength of the flamelet across the strain-rate spectrum. As shown, the integrated burning rate increases approximately linearly with strain rate, reaching its maximum just before the extinction limit. For each strain rate below extinction, two distinct flamelet solutions exist, corresponding to the upper (stable) and lower (unstable) branches of the classical S-curve. However, this multiplicity is lost when the solutions are mapped to the progress-variable space via $\lambda$. The resulting profile assumes a Gaussian-like shape, peaking at the value of $\lambda$ corresponding to the quenching-limit strain rate and vanishing at both the lower and upper bounds of the progress variable. The results reported in Fig. \ref{fig:qmapping} correspond to the flamelet solutions of Eq. (\ref{eq:steady_flamelet}), rather than resolved-scale fields.

In the FPV framework, the resolved-scale mean progress variable, $\widetilde{C}$, is typically used to document the coupling of the flamelet solutions and is therefore the quantity most often reported in figures. However, for a given combination of $\widetilde{Z}$, $\widetilde{Z''^2}$ and $\bar{p}$, the corresponding monotonically increasing curves of $\widetilde{C}_{\mathrm{tab}}(\lambda)$ must be inverted to recover the flamelet parameter $\lambda$ which determines where the local flamelet is located on the s-shaped curve, and thus, its strain rate. Figure \ref{fig:fpv_lambda} shows the spatial distribution of the normalized flamelet parameter, $\lambda/\lambda_{\mathrm{max}}$, on the resolved scale. 

Far from the mixing layer, where $\widetilde{Z}$ approaches 0 or 1, the flamelet-based profiles of $\widetilde{C}_{\mathrm{tab}}(\lambda)$ yield extremely small maximum values, since the product mass fractions at these mixture-fraction limits are asymptotically zero. Consequently, even very small numerical values of $\widetilde{C}$ that have diffused away from the reaction zone can correspond to artificially large values of $\lambda$, which explains why the air-side freestream shows a normalized flamelet parameter close to unity. For this reason, meaningful interpretation of $\lambda$ should be restricted to the vicinity of the mixing layer or reaction zone. Within this region, $\lambda$ is observed to decrease monotonically from the air side toward the fuel side across the mixing layer, with no apparent correspondence to the resolved-scale strain-rate distribution shown in Fig.~\ref{fig:sr_combined}.

Figure~\ref{fig:fpv_lines} illustrates the details of the progress-variable coupling at the streamwise location x = 25 mm. The resolved-scale progress variable $\widetilde{C}$, obtained from its transport equation, exhibits a Gaussian-like distribution centered within the flame zone, where its source term, $\widetilde{\dot{\omega}}_C$, is active. The corresponding distribution of the normalized flamelet parameter, $\lambda/\lambda_{\mathrm{max}}$, shows values approaching unity, which correspond to the equilibrium flamelet solutions associated with zero strain rate (see Fig.~\ref{fig:qmapping}). This occurs despite the fact that the local flamelet strain rate $S^*$, computed from $\epsilon$ remains finite. Consequently, in regions where $\lambda/\lambda_{\mathrm{max}} = 1$, the resolved scale effectively retrieves equilibrium flamelet states (including species mass fractions and chemical source terms) from the library, even though the local flow field is far from equilibrium.

Moreover, the heat-release rate $\widetilde{\dot{Q}}$ introduced to the resolved scale is found to be highly sensitive to small perturbations, which likely explains why most FPV formulations adopt an energy equation expressed in terms of sensible plus formation enthalpy, where the chemical heat-release term does not appear explicitly. In such cases, its influence is captured implicitly through the evolution of the species mass fractions retrieved from the flamelet tables. Nevertheless, when the progress variable (i.e., $\lambda$) is used as the flamelet coupling variable, the resulting species compositions retrieved from the tables do not correspond to the local strain-rate conditions.

These observations are critical. Although a defined relationship exists between the flamelet parameter $\lambda$ and the imposed strain rate $S^*$ at the flamelet scale, this correspondence does not carry over to the resolved scale. At the resolved level, the progress variable $\widetilde{C}$ is governed solely by its transport equation (Eq. (\ref{cequation})), and its evolution is dictated by local convection and diffusion processes rather than by a mechanistic connection to strain rate.

\subsection{Scaling of Chemical Rates}

\begin{figure}
\centering
\begin{subfigure}{.45\textwidth}
  \centering
  \includegraphics[width=1.0\textwidth]{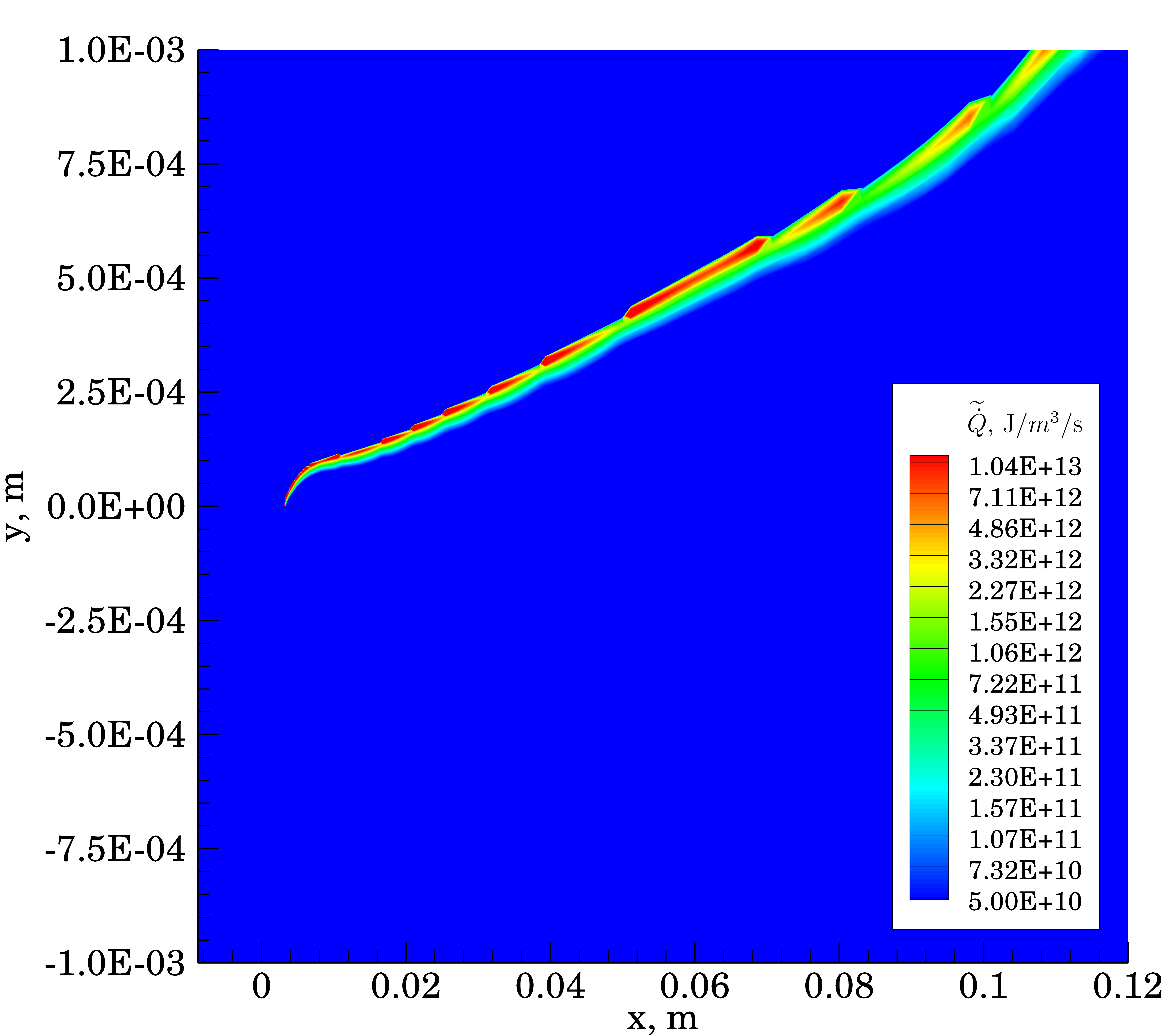}
  \caption{OSK.}
  \label{fig:osk_hrr}
\end{subfigure}%
\begin{subfigure}{.45\textwidth}
  \centering
  \includegraphics[width=1.0\textwidth]{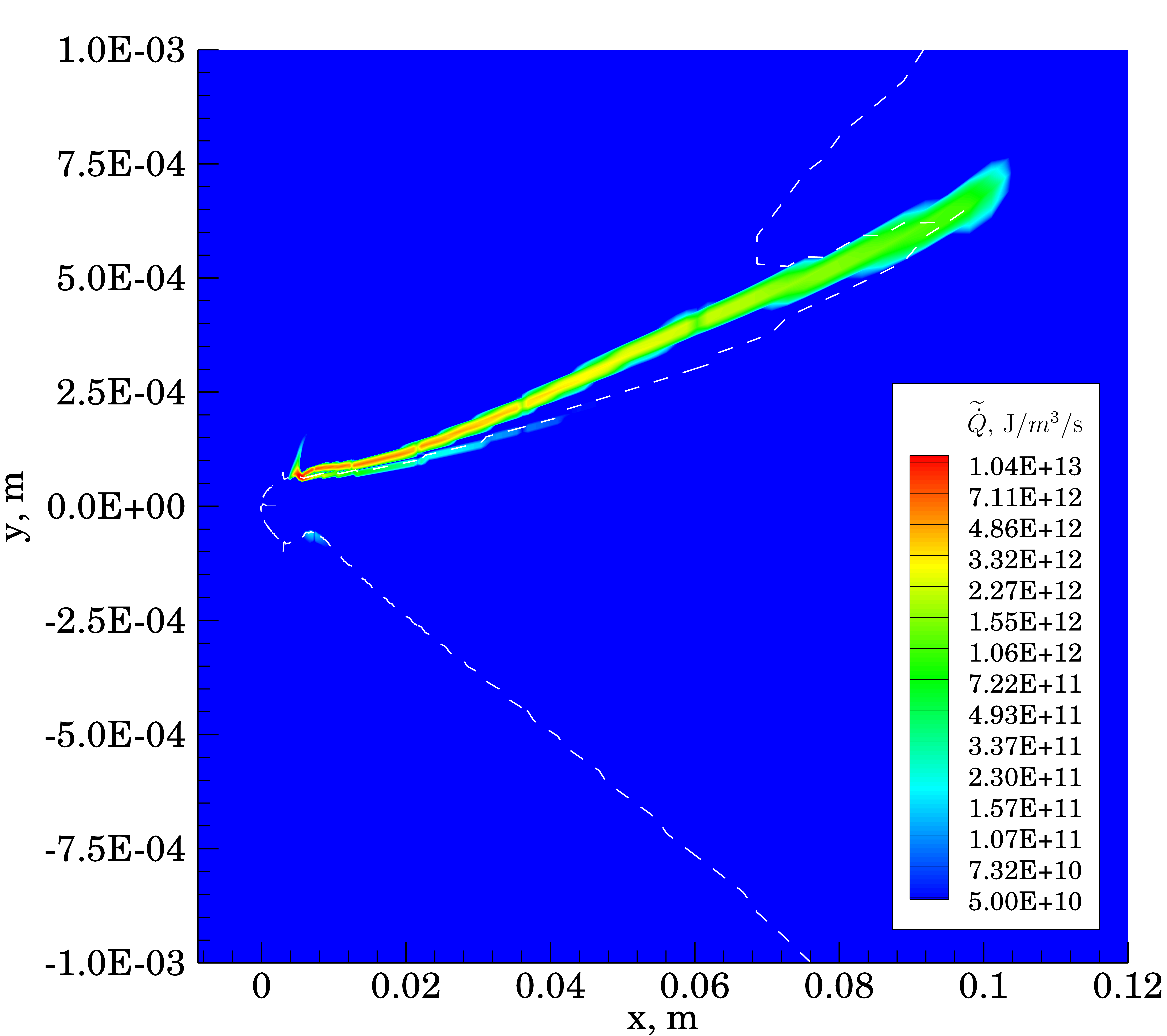}
  \caption{$\boldsymbol{\epsilon}$-based.}
  \label{fig:eps_hrr}
\end{subfigure}
\caption{Contours of heat release rate, $\boldsymbol{\widetilde{\dot{Q}}}$.}
\label{fig:heat_releases}
\end{figure}

Figure~\ref{fig:heat_releases} shows the heat-release-rate fields, $\widetilde{\dot{Q}}$, obtained from the OSK and $\epsilon$-based models. The transverse direction has been magnified to highlight the flame structure. In the 
$\epsilon$-based contour, the dashed line marks the flammability limit of the flamelet, which depends on both the local strain rate, $S^*$, and the pressure. Upstream of this line, $S^*$ remains below the flammability limit, allowing the flamelet to exist in a burning state. Downstream, 
$S^*$ exceeds the limit and the flamelet becomes quenched. This behavior is clearly visible immediately downstream of the splitter plate, where the high shear leads to local quenching and the flame standoff observed in Fig.~\ref{fig:tempcont}. A second quenching region appears near x = 100 mm corresponding to the decrease in flammability limit with pressure, consistent with the temperature field shown in Fig.~\ref{fig:tempcont}. Relative to the OSK formulation, the $\epsilon$-based model shows a faster decay in peak heat release along the reaction zone. This behavior is attributed to the stronger dependence of the detailed finite-rate kinetics on local pressure, which is explicitly represented in the $\epsilon$-based flamelet tables.

Both the OSK and $\epsilon$-based models predict comparable reaction-zone widths at the resolved scale, which explains the similar thermal mixing-layer thicknesses reported in Fig.~\ref{fig:temp_lines}. When compared with the FPV results in Fig.~\ref{fig:fpv_lines}, the 
$\epsilon$-based model produces a much thinner reaction zone. This occurs because the $\epsilon$-based approach includes resolved-scale species transport equations, rather than relying solely on the mixture-fraction distribution (as in the FPV model), to determine the spatial distribution of the chemical rates. 

In the $\epsilon$-based framework, the local resolved-scale species concentrations directly influence where chemical reactions can occur. The chemical source terms $\widetilde{\dot{\omega}}_n$ obtained from the flamelet library are therefore scaled according to the availability of reactants on the resolved scale. Specifically, for each species $n$, the source term is corrected as
\begin{equation}
\widetilde{\dot{\omega}}_{n_{\mathrm{corr}}} = \alpha\widetilde{\dot{\omega}}_n, \qquad n = 1, 2, \ldots, N,
\end{equation}
where the scalar coefficient $\alpha$ is chosen independently at each computational cell as the minimum value that ensures positivity of all species mass fractions. This correction prevents unphysical behavior that could occur when a flamelet, evaluated at a given $\widetilde{Z}$, predicts finite or negative production rates for reactants whose resolved-scale mass fractions are zero or vanishingly small. In effect, the scaling enforces that no chemical source term is applied in regions devoid of reactants, thereby ensuring a physically consistent coupling between the flamelet chemistry and the resolved species fields.
Admittedly, the above rate-scaling procedure is somewhat ad hoc. However, it becomes necessary because the flamelet libraries are generated under fixed scalar boundary conditions for temperature and species mass fractions, corresponding to the inflow values of the resolved-scale reactant streams. This assumption is standard practice \cite{nguyen_longitudinal_2018,pecnik_reynolds-averaged_2012,saghafian_efficient_2015,nguyen_driving_2017,nguyen_impacts_2018,nguyen_spontaneous_2019,shadram_neural_2021,shadram_physics-aware_2022,zhan_combustion_2024,shan_improved_2021,jiang_species-weighted_2023,ihme_regularization_2012,najafi-yazdi_systematic_2012,bojko_formulation_2016} and is also employed in conventional FPV formulations.

A natural next step would be to extend the flamelet database to include varying scalar boundary conditions, thereby accounting for different local combinations of reactant concentrations and temperatures. Such an approach would eliminate the need for post-scaling of the chemical source terms, as the flamelet state could be directly retrieved for each relevant local boundary condition. However, this enhancement would entail a significant increase in computational cost and data dimensionality for the flamelet library generation since a significantly increased  number of additional degrees of freedom must be tabulated.

\begin{figure}
\centering
\begin{subfigure}{.5\textwidth}
  \centering
  \includegraphics[width=1.0\textwidth]{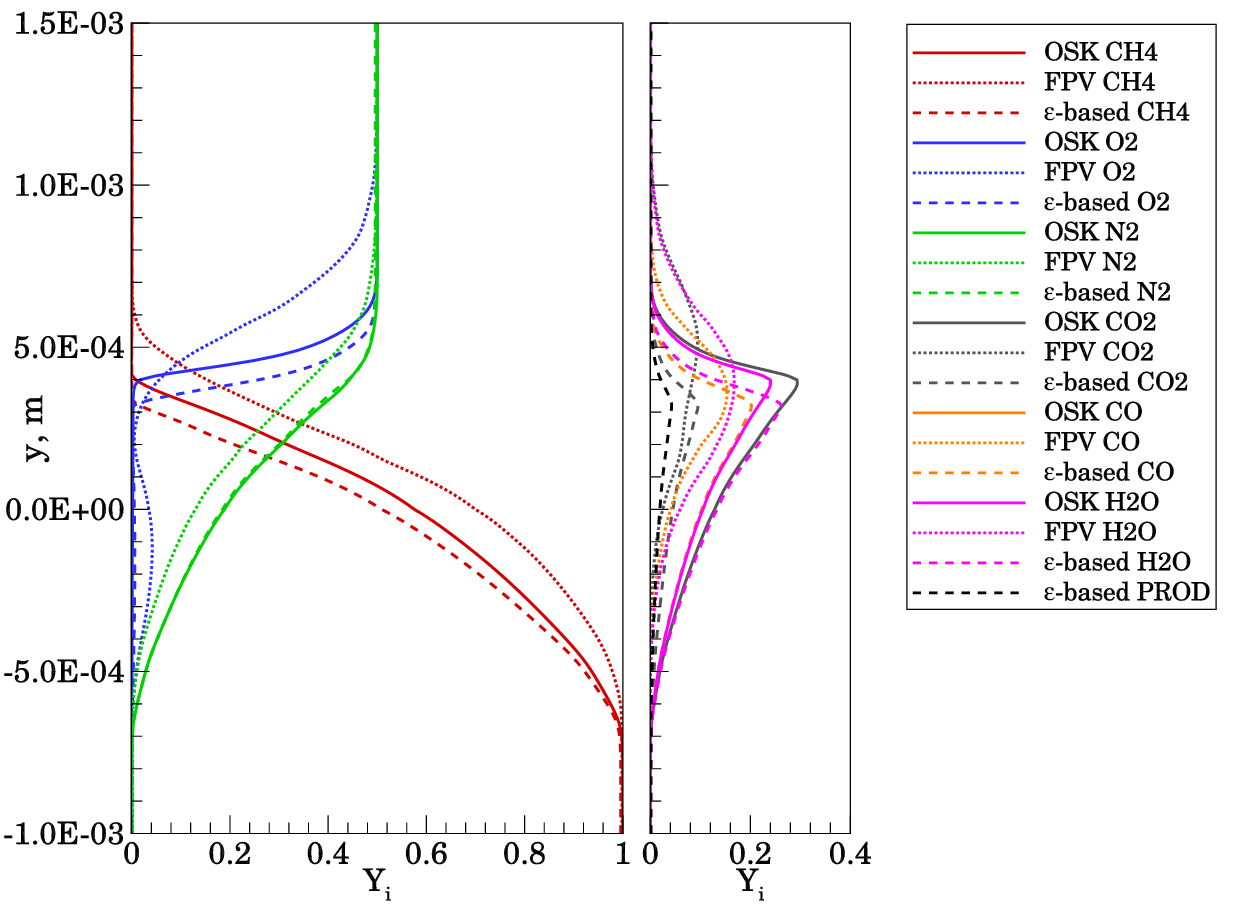}
  \caption{Species profiles at x = 50 mm.}
  \label{fig:y_profiles}
\end{subfigure}%
\begin{subfigure}{.5\textwidth}
  \centering
  \includegraphics[width=0.8\textwidth]{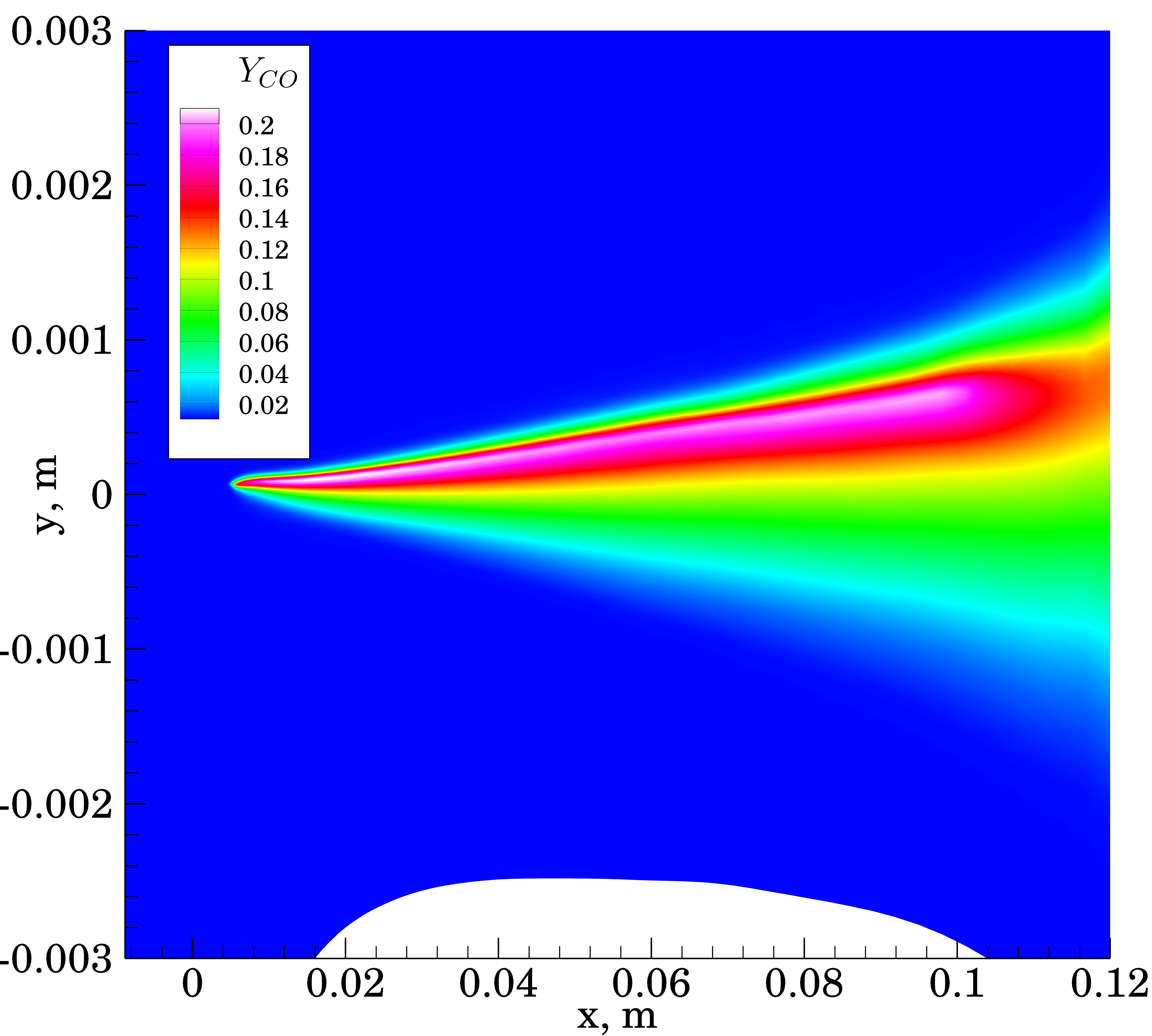}
  \caption{Contour of $\boldsymbol{\widetilde{Y}_{\mathrm{CO}}}$ for the $\boldsymbol{\epsilon}$-based results.}
  \label{fig:y_co}
\end{subfigure}
\caption{Mixture compositions and $\boldsymbol{\widetilde{Y}_{\mathrm{CO}}}$ field.}
\label{fig:mass_fractions}
\end{figure}

Accounting for species transport on the resolved scale through explicit species transport equations, as done in the OSK and $\epsilon$-based models, produces noticeably different mixture compositions compared to retrieving species compositions from the precomputed tables as in the FPV approach. Figure~\ref{fig:y_profiles} shows the mixture composition profiles at $x = 50$ mm for all three models. The FPV predicts broader profiles, consistent with its wider thermal and velocity mixing layers, but these are also more diffusive. Furthermore, FPV yields nearly symmetric product distributions, whereas the OSK and $\epsilon$-based results are skewed toward the fuel side, consistent with the thicker fuel-side mixing layer. 

Another FPV artifact is the entrainment of $\mathrm{O_2}$ into the fuel side, producing a local $\mathrm{O_2}$ maximum below the centerline. The OSK does not exhibit this behavior because ignition occurs immediately at the splitter-plate trailing edge. In the $\epsilon$-based model, some $\mathrm{O_2}$ entrainment occurs due to the flame standoff region, where oxidizer is advected into the fuel side, but the local maximum decays downstream as turbulent diffusion acts on the gradients. These differences arise because, in the FPV model, transport of species acts indirectly through the transport  of $\widetilde{Z}$, $\widetilde{Z''^2}$, and $\widetilde{C}$, rather than directly on species gradients. Hence, explicitly solving the species transport equations may better capture transport effects at the resolved scale, which could justify the added computational cost associated with solving resolved-scale species transport. Another feature in these profiles is the lumped species “PROD” in the $\epsilon$-based model, whose resolved-scale mass fraction remains bounded by the subgrid constraint $\gamma \ge 0.95$ (Eq.~\ref{eq:prod}), i.e., $Y_{\mathrm{PROD}} \le 5\%$.

Figure~\ref{fig:y_co} shows contours of $Y_{\mathrm{CO}}$ for the $\epsilon$-based model. Despite local flamelet quenching at $x = 100$ mm, where reaction rates drop to zero, the resolved-scale species transport equations allow CO produced upstream to advect and diffuse downstream of the quenched region. This behavior mitigates a key limitation of strain-rate-based formulations, in which direct interpolation of species mass fractions from flamelet tables leads to abrupt transitions between burning and non-burning states along the S-curve. The FPV model introduces the progress variable specifically to smooth these discontinuities by providing a continuous mapping between flamelet branches. However, our results show that when species transport is explicitly solved on the resolved scale, the species concentrations remain continuous across the quenched region without the need for a progress variable, while only the chemical source terms exhibit discontinuities.

\section{Conclusions}

Two-dimensional RANS simulations were performed to evaluate the performance of the conventional FPV model and a novel $\epsilon$-based flamelet model for diffusion-flame configurations. The analysis focused on the physical consistency of coupling between resolved-scale and subgrid quantities, particularly the role of strain rate in determining the local flamelet state.

The FPV results show that decoupling between resolved-scale and flamelet-scale strain rates in the progress-variable formulation leads to preferential selection of equilibrium flamelet solutions in regions of high strain. This behavior produces nonphysical predictions of heat release rates and species mass fractions in such regions.

To address this limitation, a correction was introduced in which the turbulent kinetic energy dissipation rate, $\epsilon$, replaces the progress variable as the tracking parameter, allowing the flamelet solutions to be parameterized by the inflow strain rate, $S^*$. The strain rate $S^*$, computed using a gradient-based scaling argument with $\epsilon$, exhibits strong spatial correlation with the resolved-scale strain-rate field. As a result, when $\epsilon$ is used as the flamelet tracking variable instead of the progress variable, the subgrid flamelet responds consistently to the resolved-scale strain field.

Using strain rate as an input to the flamelet tables enables the prediction of key resolved-scale flame features, such as flame standoff and local quenching, which reflect a direct response of the flamelet to the resolved-scale strain field. This formulation removes the FPV model’s sensitivity to the specific definition of the progress variable.

In addition, the inclusion of resolved-scale species transport in the $\epsilon$-based model allows advective and diffusive redistribution of products across locally quenched regions, thereby eliminating discontinuities that arise when the reaction field transitions between burning and non-burning states. This relaxes the FPV model’s long-standing limitation associated with unphysical ignition and extinction jumps along the S-curve.

Although the $\epsilon$-based model introduces additional transport equations compared with the FPV model, its computational cost remains far below that of fully resolved finite-rate chemistry simulations. Furthermore, the number of tracked resolved-scale species can be reduced using lumping strategies.

Overall, the results demonstrate that $\epsilon$ can serve as a physically consistent coupling variable between resolved-scale quantities and subgrid flamelet dynamics, offering potential improvements in the representation of extinction behavior and strain-rate response. Future work should extend this framework to three-dimensional LES configurations and assess its performance through experimental validation.


\section*{Acknowledgments}
The research was supported by the Office of Naval Research through Grant N00014-22-1-2467 with Dr. Steven Martens as program manager. Professor Heinz Pitsch of RWTH Aachen University is acknowledged for providing us access to the FlameMaster code. Artificial intelligence tools were used solely to improve grammar and language clarity during manuscript preparation. All technical content, figures, and analyses are original.

\bibliography{references}

\end{document}